%% file: ms.tex
\documentclass[twocolumn,onecolappendix]{aastex62}
\usepackage{amsmath}

\shortauthors{Sheehan et al.}
\shorttitle{Substructures in Protostellar Disks}

\begin{document}

\title{\bf The VLA/ALMA Nascent Disk and Multiplicity (VANDAM) Survey of Orion Protostars. III. Substructures in Protostellar Disks}

\author{Patrick D. Sheehan}
\affiliation{Center for Interdisciplinary Exploration and Research in Astronomy, 1800 Sherman Rd., Evanston, IL 60202, USA}
\affiliation{National Radio Astronomy Observatory, 520 Edgemont Rd., Charlottesville, VA 22901, USA}
\affiliation{Homer L. Dodge Department of Physics and Astronomy, University of Oklahoma, 440 W. Brooks Street, Norman, OK 73019, USA}
\affiliation{NSF Astronomy \& Astrophysics Fellow}

\author{John J. Tobin}
\affiliation{National Radio Astronomy Observatory, 520 Edgemont Rd., Charlottesville, VA 22901, USA}
\affiliation{Homer L. Dodge Department of Physics and Astronomy, University of Oklahoma, 440 W. Brooks Street, Norman, OK 73019, USA}

\author{Sam Federman}
\affiliation{Ritter Astrophsical Research Center, Department of Physics and Astronomy, University of Toledo, Toledo, OH 43606, USA}

\author{S. Thomas Megeath}
\affiliation{Ritter Astrophsical Research Center, Department of Physics and Astronomy, University of Toledo, Toledo, OH 43606, USA}

\author{Leslie W. Looney}
\affiliation{Department of Astronomy, University of Illinois, Urbana, IL 61801, USA}

\begin{abstract}
The prevalence of substructures in $\sim1-10$ Myr old protoplanetary disks, which are often linked to planet formation, has raised the question of how early such features form, and as a corollary, how early planet formation begins. Here we present observations of seven protostellar disks (aged $\sim0.1-1$ Myr) from the VLA/ALMA Nascent Disk and Multiplicity Survey of Orion Protostars (VANDAM: Orion) that show clear substructures, thereby demonstrating that these features can form early in the lifetimes of disks. We use simple analytic models as well as detailed radiative transfer modeling to characterize their structure. In particular we show that at least four of the sources have relatively massive envelopes, indicating that they are particularly young, likely the youngest disks with substructures known to-date. Several of these disks also have emission from an inner disk that is offset from the center of the ring structure. Given the size of the cleared out regions of the disk, it is unclear, however, whether these features are related to planet formation, or rather if they are signposts of close-separation binary formation at early times.
\end{abstract}


\section{Introduction}

High resolution imaging with the Atacama Large Millimeter/Submillimeter Array (ALMA) in recent years has shown that protoplanetary disks are not smooth and featureless, but rather that substructures appear to be ubiquitous in disks \citep[e.g.][]{Brogan2015,Andrews2016,Andrews2018,Long2018}. The most frequently found forms of such substructures are azimuthally symmetric bright and dark rings \citep[e.g.][]{Huang2018,Long2018} that are often associated with a depletion of dusty material in the the dark rings, hence their commonly being referred to as gaps. Azimuthally asymmetric features like spirals \citep[e.g.][]{Perez2016,Huang2018b} and vortices \citep[e.g.][]{vanderMarel2013} are also found in many disks. The presence of such features likely provides a solution to the long-standing radial drift problem: large dust grains are expected to lose angular momentum due to a gas headwind and spiral into the central protostar on timescales much shorter than the lifetime of the disk \citep[e.g.][]{Weidenschilling1977b}. The presence of these substructures suggests that pressure bumps in the disk trap large particles and prevent them from drifting inwards any further \citep[][]{Pinilla2012}.

Though pressure bumps may solve the radial drift problem and allow for long-lived protoplanetary disks, they also raise other questions. Two questions of particular note are: 1. How do these substructures form? And, 2. How early do they form? The former has been explored in great detail, and a number of mechanisms that are capable of forming substructures have been proposed: disk photoevaporation \citep[e.g.][]{Alexander2006,Gorti2009a,Owen2010}, dust grain growth \citep[e.g.][]{Dullemond2005,Birnstiel2012}, processes driven by the presence of snowlines \citep[e.g.][]{Clarke2001,Zhang2015,Okuzumi2016}, magnetic zonal flows \citep[e.g.][]{Flock2015}, and perhaps the most popular mechanism, dynamical sculpting by a large body, possibly of planetary mass \citep[e.g.][]{DodsonRobinson2011,Dong2015}. 

\input{table1.tex}

There is also tantalizing evidence that substructures are formed early in the lifetimes of protoplanetary disks. HL Tau, the first protoplanetary disk found to have narrow bright and dark rings \citep{Brogan2015}, is thought to be a relatively young star that is still embedded in some remnant envelope material, left from its initial gravitational collapse \citep[often classified as a ``Flat Spectrum" protostar; e.g.][]{Furlan2008}. Substructures such as azimuthally symmetric bright and dark rings as well as cavities stretching all the way to the central star have also been found in the disks of several sources classified as embedded protostars \citep[WL 17, GY 91, and DG Tau B; also known as Class 0, I \& Flat Spectrum protostars, and ``protostellar" disks;][]{Sheehan2017a,Sheehan2018,deValon2020}, which are typically thought to have ages $\lesssim0.5$ Myr \citep[e.g.][]{Evans2009,Dunham2015}. However, careful analysis of those particular sources has suggested that they may be late-stage protostars, with relatively little envelope material left over \citep[e.g.][]{Sheehan2017a,Sheehan2018}. Clear spiral arms have also been found in L1448IRS3B and HH111VLA1, both embedded sources, that are likely driven by gravitational instabilities in particularly massive disks \citep[e.g.][Reynolds et al., in prep.]{Tobin2016b,Lee2019b}.

The substructures found in protostellar disks thus far have come from one-off observations from an in-homogeneous sample. However, recent millimeter surveys disks have begun to collect resolved ALMA observations of large samples of protostellar disks in order to study their structures, including any substructures that may be present. Here we present the seven protostellar disks from the VLA/ALMA Nascent Disk and Multiplicity Survey of Orion Protostars (VANDAM: Orion), the most comprehensive of such surveys, that show substructure at $\sim40$ au spatial resolution. This sample doubles the number of protostellar disks currently known to have substructures, and presents an opportunity to begin to characterize these features early ($\lesssim1$ Myr) in the lifetimes of disks. 

The structure of this work is as follows: In Section \ref{section:observations} we discuss the VANDAM: Orion survey and how our sample of disks was selected. In Section \ref{section:analysis} we fit both simple geometrical models as well as detailed radiative transfer models to our data to characterize these disks and their substructures. And finally in Section \ref{section:discussion} we discuss the implications of finding these substructures in protostellar disks and what their origins may be.

\section{Sample Selection \& Data Reduction}
\label{section:observations}

\begin{figure*}[t]
\centering
\includegraphics[width=7in]{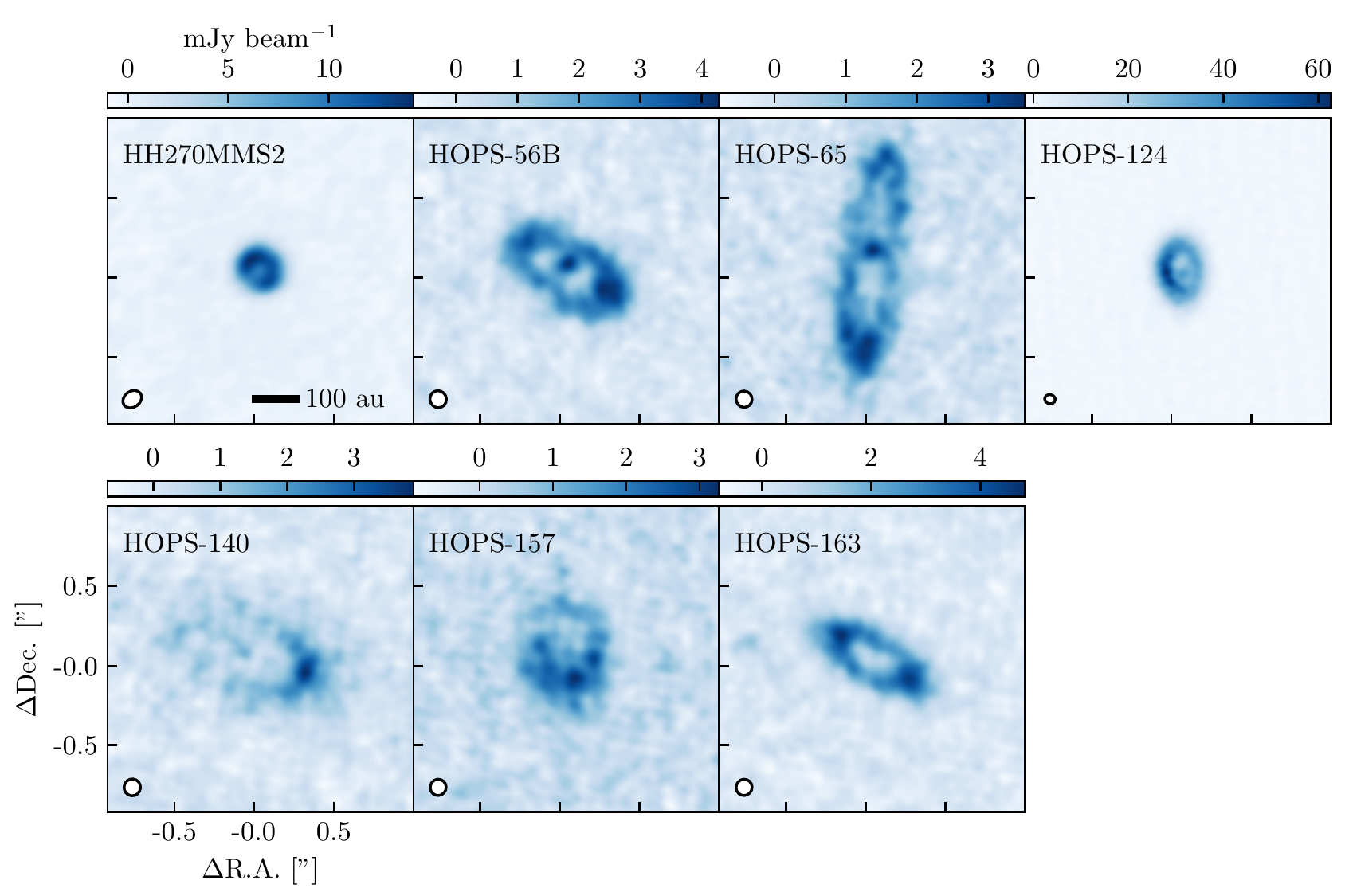}
\caption{ALMA 345GHz dust continuum images of the seven disks with clear gaps/cavities from the VANDAM: Orion sample. A line with a length corresponding to 100 au is shown in the top left panel, for scale.}
\label{fig:td_images}
\end{figure*}

The VANDAM: Orion Survey targeted $>300$ embedded (Class 0, I, \& Flat Spectrum) protostars in the Orion Molecular Cloud Complex \citep[$d \sim 400$ pc;][]{Kounkel2017,Kounkel2018,Zucker2019} with ALMA at 345 GHz and $\sim$0.1" spatial resolution ($\sim40$ au at the distance of Orion; Project code: 2015.1.00041.S; for further details see \citet{Tobin2020}. Details of the data reduction, self-calibration when the signal-to-noise ratio was high enough, and imaging can be found in the survey overview paper \citep{Tobin2020}. We note that self-calibration was done for all of the sources considered in this manuscript.

From this sample we selected every source that, by visual inspection, showed clear and unambiguous evidence of annular substructures/local extrema (commonly referred to as "rings" and "gaps") in its disk, amounting to a total of 7 disks: HOPS-56B, HOPS-65, HOPS-124, HOPS-140, HOPS-157, HOPS-163, and HH270-MMS2. For most of our sources this inspection was done in images generated using Briggs weighting with a robust parameter of 0.5. We did, however, generate images for each source in the VANDAM: Orion survey using a range of robust parameters as well as superuniform weighting, and images made with higher spatial resolution did often help to find new sources or confirm hints of substructure found in lower resolution images. A few additional sources with tentative hints of substructure were also considered (HOPS-80, HOPS-102, HOPS-250, HOPS-260, and HOPS-368). Ultimately, though the images were tantalizing, we excluded these sources out of caution to not falsely identify imaging artefacts as substructures. 

We report the source properties from \citet{Tobin2020} for all 7 disks in our sample in Table \ref{table:source_properties}. The Gaussian fitting that \citet{Tobin2020} used to determine source properties may be inaccurate for these sources with substructures because their resolved structure cannot be reasonably described by a single-component Gaussian. Instead, we use the total flux of the disk-associated components of our analytic modeling (see Section \ref{section:analytic_models}) in place of the \citet{Tobin2020} 0.87 mm fluxes, as the previously done Gaussian fitting is primarily sensitive to the disk-like structures in the images. We re-calculate the relevant spectral indices, as well, using these updated fluxes. 

We show images of our sample in Figure \ref{fig:td_images}. All of these images were made with a robust parameter of 0.5 and a beam size of $\sim0.1"$, with the exception of HOPS-124, which has high signal-to-noise and is quite compact, so we used superuniform weighting to achieve a resolution of 0.06".

For most sources we use the ALMA visibilities from the (self-)calibrated Measurement Set files, without modification, for our analysis. A few of our targets, though, are either multiples (HOPS-56B), or have nearby protostars within their fields of view (HOPS-140). For these sources, to isolate the visibility data for the sources that are relevant for this paper, we first run the CASA {\it clean} task to generate a $clean$-component model of the data. We then mask out the $clean$-components associated with the source of-interest, and subtract the remaining $clean$ components from the data in the visibility plane using the CASA $uvsub$ task. The visibility data that remains after this procedure should well represent the visibilities for the source of interest, and we confirm this by re-imaging the data to ensure that there are no traces remaining of the additional sources and that the structures of the sources of interest are not altered.

In addition to the VANDAM: Orion ALMA observations, a number of our sources (HOPS-65, HOPS-124, HOPS-140, HOPS-157, and HOPS-163) were observed as part of program code 2018.1.01284.S (PI: Megeath) with the ALMA Compact Array 7M array (ACA), and those observations are used to compare with our radiative transfer modeling (see Section \ref{section:rt_results}). These observations were conducted between 2 October and 30 October 2018. The spectral windows for the continuum imaging were centered on 332.975 and 343.975 GHz, with the correlator providing bandwidths of 2.000 GHz at each frequency. On-source integration times ranged from 4500 to 6000 seconds, and were based on the 870 um single dish flux. The minimum baseline was 8.9 m, and the maximum was 48.9 m. The reduction of the ACA 870 micron data was executed following the standard imaging procedures provided by CASA. Calibration was completed using the basic reduction scripts included in the data package for running the ALMA pipeline. The fluxes measured from these observations are listed in Table \ref{table:source_properties}. The discrepancy between the ACA and main array fluxes apparent for HOPS-140 is likely due to flux calibration uncertainties, which can also be seen by the offset between the two in the azimuthally averaged visibility plot for HOPS-140 in Section \ref{section:rt_results}.

We note that though we show these ACA data for comparison with our radiative transfer modeling fits, the data themselves were not included in the fitting procedure (for either the analytic, or radiative transfer modeling), as they were unavailable until after much of that work had been done, and the models are computationally expensive to re-run. Nonetheless, they provide an important comparison for the radiative transfer models to ensure that we are recovering large scale structure properly.

Finally, as much of this work relies on fitting our data with models, we compare the weights from the calibrated ALMA visibilities with the root mean square (RMS) of naturally weighted images by estimating the uncertainties from combined visibilities as $\sigma_{vis} = \sqrt{\sum{1/w_i}}$, where $w_i$ is the weight of the ith visibility. We find that on average, the weights provided by the ALMA pipeline need to be multiplied by a factor of $\sim0.25$ to match the RMS of the images, so for any model fitting we do we have adjusted the weights by this factor. Though this may not be a perfect comparison, the scaling we use provides a conservative estimate of the uncertainties on the data.

In addition to the ALMA millimeter observations, we collect supplementary photometric and spectroscopic data from the {\it Herschel} Orion Protostar Survey \citep[HOPS;][]{Fischer2013}. This dataset includes near-infrared Spitzer IRAC and MIPS photometry, Herschel PACS photometry, 2MASS J, H, and K photometry, and Spitzer IRS spectroscopy. When analyzing the spectroscopic data (see Section \ref{section:rt_models}) we bin the IRS spectra into 25 points evenly spaced throughout the wavelength coverage to avoid over-weighting the IRS spectrum compared with the broadband photometry, as well as costly radiative transfer calculations at hundreds of wavelengths. We assume a uniform, 10\% flux calibration error on all photometry, including the binned IRS spectra, for the purposes of fitting models to these data.

\section{Analysis \& Results}
\label{section:analysis}

The images in Figure \ref{fig:td_images}, of our sample of 7 protostellar disks, show a range of interesting substructures similar to what has been found in protoplanetary disk samples, typically at closer distances. Each of the disks in our sample is dominated by a single bright ring. For five of those sources (HOPS-56B, HOPS-65, HOPS-124, HOPS-140, and HOPS-157), that bright ring has a clear brightness asymmetry. Furthermore, four disks in our sample (HOPS-56B, HOPS-65, HOPS-124, and HOPS-140) have evidence of point source-like features interior to the bright ring, indicating the presence of an inner disk.

\input{table2.tex}

All of these protostars have been classified as Class 0, I, or Flat Spectrum, indicating that they are young, and embedded within an envelope. These disks are among the youngest disks found to have substructure. As such, understanding their structures is likely to illuminate our understanding of how such features develop, and how disks evolve. To better characterize and understand these features, we employ two separate modeling approaches: 1) First, we fit a combination of simple analytic functions designed to fully represent the intensity profile of the disks. These models are designed to fully characterize the geometrical properties of the disk intensity structure, but are not based on an underlying physical model. 2) Second, we fit physically motivated radiative transfer models to our data. These are designed, in particular, to gain a better picture of the physical structure of these systems to better understand their evolutionary status, beyond simplistic classifications like the infrared spectral index that is often used. We describe each of the modeling procedures, and their results, below.

\subsection{Analytic Models}
\label{section:analytic_models}

\begin{figure*}
    \centering
    \includegraphics[width=7in]{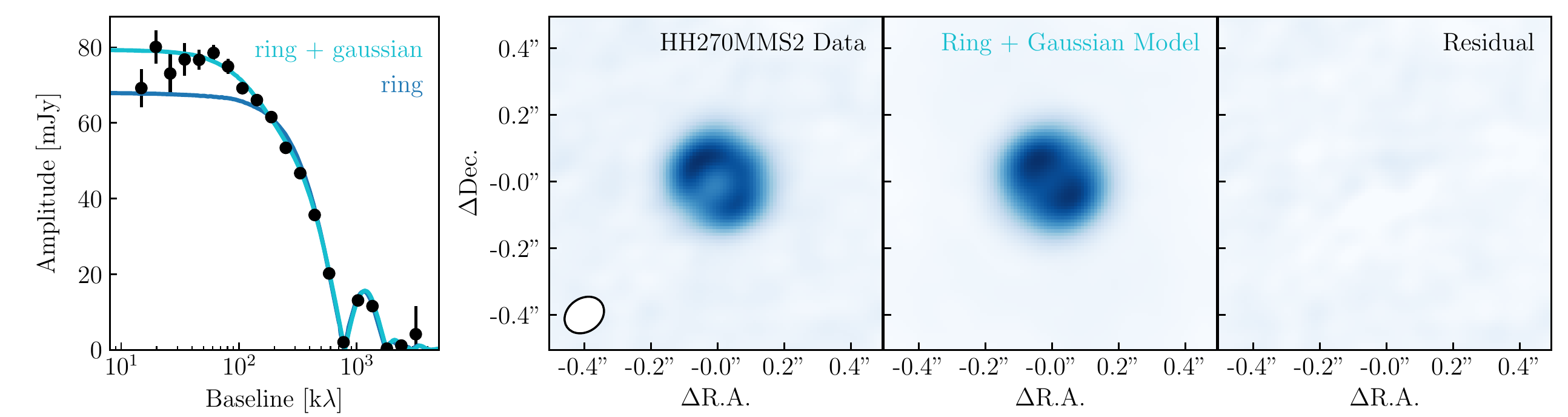}
    \includegraphics[width=7in]{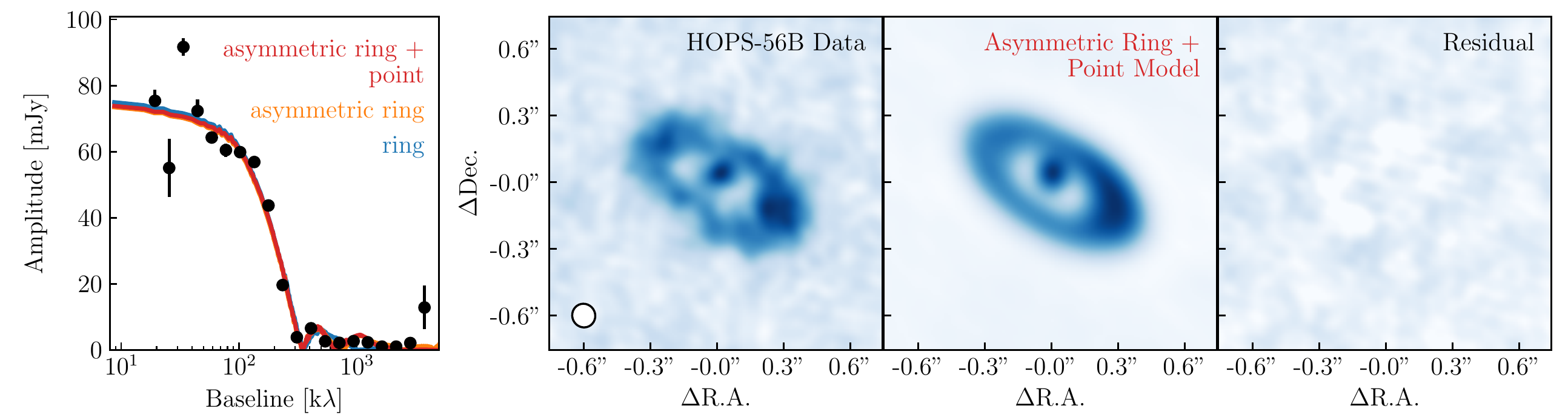}
    \caption{Examples of our analytic model fits to two of the sources in our sample. The left column shows the one-dimensional azimuthally averaged visibilities compared with the best fit model curve for different combinations of the component described in Section \ref{section:analytic_models}. The data were binned into 20 points in logarithmic space, which provided a good balance between showing features present in the data and the signal-to-noise of the averaged data, though other numbers of bins were considered to ensure consistency. Though the visibilities are shown averaged radially for ease of viewing, all fits were done to the full two dimensional data. In the latter three columns we show the images of our data, the best-fit model, and the residuals. The model and residual images were generated by Fourier transforming a model image, sampling at the same baselines as the data in the $uv$-plane using \texttt{GALARIO}, subtracting these synthetic visibilities from the data in the case of the residuals, and re-imaging with a CLEAN implementation built into \texttt{pdspy}.}
    \label{fig:analytic_fits1}
\end{figure*}

To characterize the structure of our sample of 7 protostellar disks, we begin by fitting simple analytical models to our data, motivated by the models fit in \citet{vanderMarel2015}. We use a ring model as the base, with a brightness distribution given by,
\begin{equation}
    I_{\nu} = I_{\nu,r} \exp{\left(-\frac{(r - r_{c,r})^4}{2 \, {r_{w,r}}^4}\right)},
\end{equation}
where $r_{c,r}$ is the radius of the center of the ring, and $r_{w,r}$ is the half-width of the ring. The coordinate system is specified in the image plane, with $x$ in the east-west direction and $y$ in the north-south direction. For a ring with some inclination ($i$) and position angle ($p.a.$), then the coordinate system of the ring is given by,
\begin{gather}
    x' = x \cos(p.a.) + y \sin(p.a.), \\
    y' = x \sin(p.a.) + y \cos(p.a.), \\
    r = \sqrt{{x'}^2 + \frac{{y'}^2}{\cos^2 i}}, \\
    \phi = \arctan{\frac{y'}{x' \, \cos i}}.
\end{gather}
Rather than using the peak surface brightness ($I_{\nu,r}$) as a free parameter, we use the integrated total flux of the ring component ($F_{\nu,r}$).

To this base model we add some combination, as appropriate for each individual source, of an azimuthal asymmetry component, a point source component to represent the central point-like emission found in the images of several sources, and a large scale Gaussian component to represent large scale emission from the envelope. The azimuthally asymmetric component is described by,
\begin{equation}
    I_{\nu} = I_{\nu,a} \exp{\left(-\frac{(r - r_{c,a})^4}{2 \, {r_{w,a}}^4}\right)} \, \exp{\left(-\frac{(\phi - \phi_{c,a})^4}{2 \, {\phi_{w,a}}^4}\right)}.
\end{equation}
For flexibility, we do not require that the asymmetry and the ring components be aligned ($r_{c,r} \neq r_{c,a}$), nor that their widths be the same ($r_{w,r} \neq r_{w,a}$). $\phi_{c,a}$ is relative to the major axis of the disk with positive values in the counter-clockwise direction. 

The point source component is given by,
\begin{equation}
    I_{\nu} = I_{\nu,p} \exp{\left(-\frac{(x' - x_c)^4 + (y' - y_c)^4}{2 \, {r_{w,p}}^4}\right)}
\end{equation}
with
\begin{gather}
    x_c = r_{c,p} \, \cos{\phi_{c,p}} \\
    y_c = r_{c,p} \, \sin{\phi_{c,p}} \, \cos{i},
\end{gather}
with $\phi_{c,p}$ once again relative to the major axis of the disk, with positive values in the counterclockwise direction. We require that the point source component falls within the inner edge of the ring and asymmetric structure.

\begin{figure*}
    \centering
    \includegraphics[width=7in]{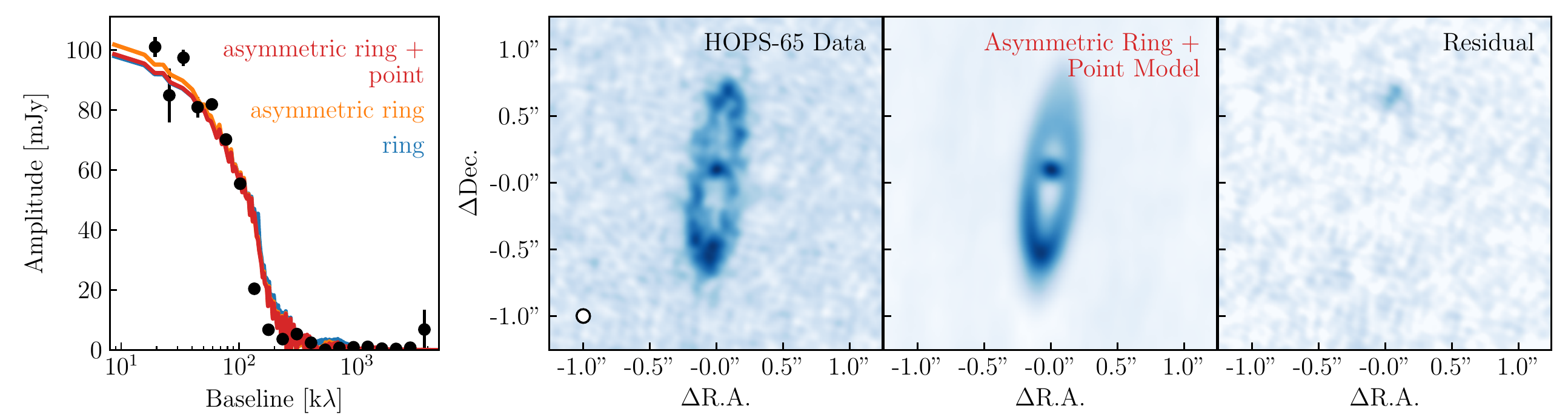}
    \includegraphics[width=7in]{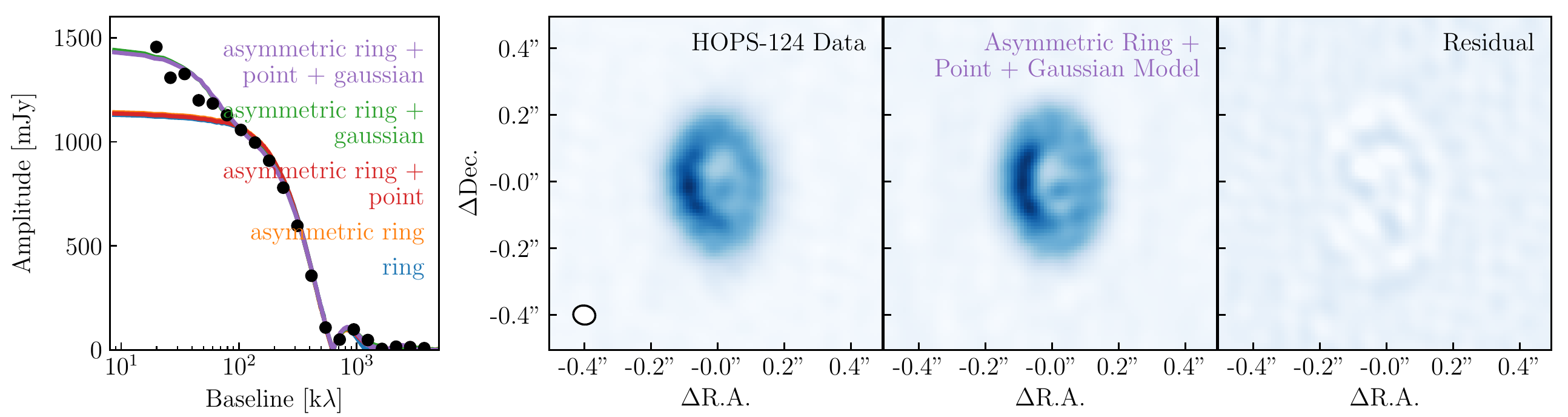}
    \includegraphics[width=7in]{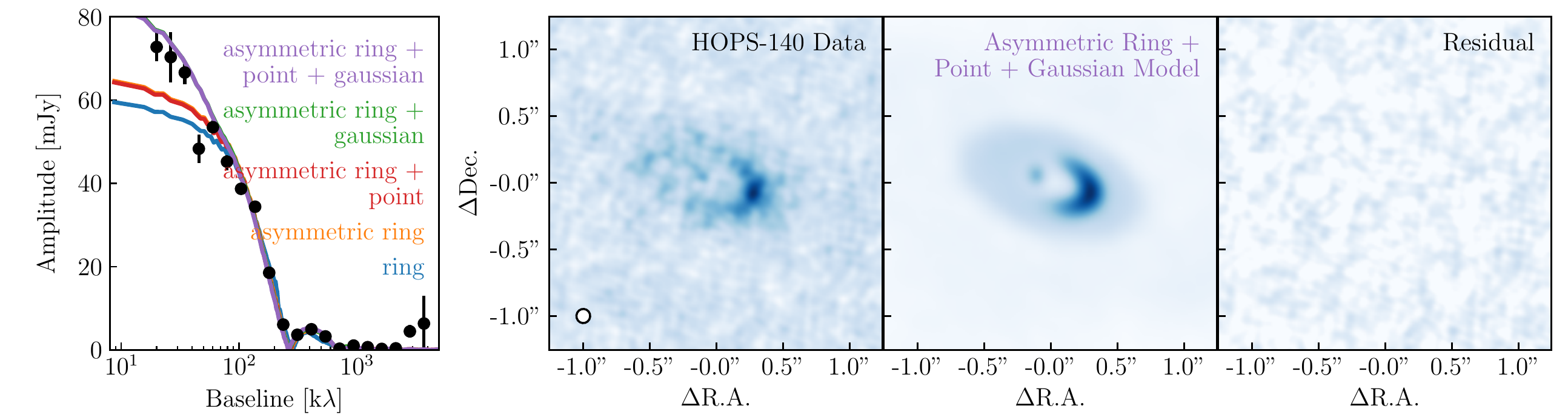}
    \includegraphics[width=7in]{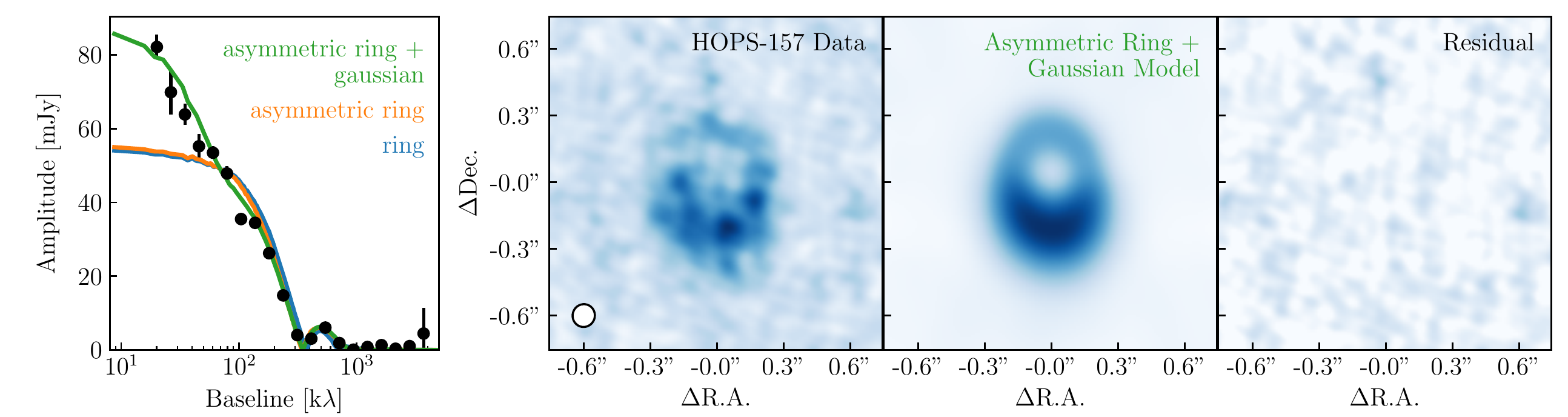}
    \includegraphics[width=7in]{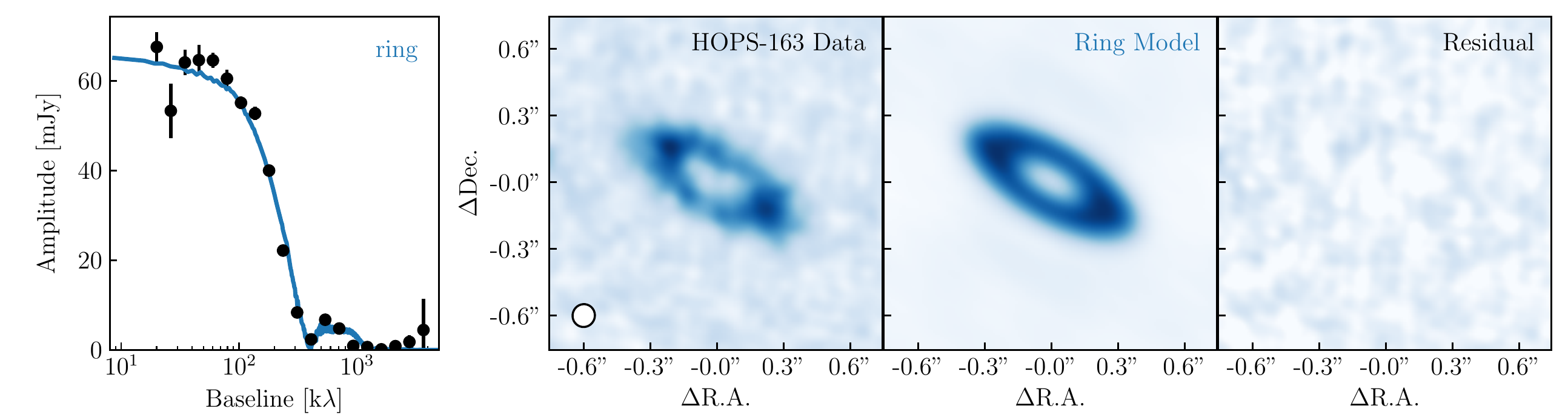}
    \caption{A continuation of Figure \ref{fig:analytic_fits1} showing the best-fit analytic models compared with the data for the remaining five sources in our sample.}
    \label{fig:analytic_fits2}
\end{figure*}

Finally, the large scale Gaussian component is given by,
\begin{equation}
    I_{\nu} = I_{\nu,G} \exp{\left(-\frac{x^2 + y^2}{2 \, {r_{w,G}}^2}\right)}.
\end{equation}
The full model is then the addition of all appropriate components for a given source,
\begin{equation}
    I_{\nu} = I_{\nu,r} + I_{\nu,a} + I_{\nu,p} + I_{\nu,G}.
\end{equation}
For each of the components, as was the case with the base ring component, rather then using the peak surface brightness as a free parameter, we use the total integrated flux ($F_{\nu,*}$). We also require that the flux of the asymmetry component be smaller than the flux of the ring component, and that the flux of the point source component be smaller than the flux of the asymmetry component. Though the final requirement may not necessarily need to be the case for all sources, it appears to be valid here, and helps the fits avoid converging to certain pathological cases.

We fit this model directly to our ALMA visibility data (not including the ACA observations) in the $uv$ plane. Rather than analytically Fourier transforming the intensity profile of our model, we generate an image-plane model image, and use the \texttt{GALARIO} package \citep{Tazzari2018} to compute the Fourier transform of that image sampled at the same baselines as our observations. We also apply a source offset from phase center to the visibility data, adding two additional parameters ($x_0$, $y_0$). In total, there are up to 18 parameters that can be included in the model, $\hat{\theta} = \{r_{c,r}, r_{w,r}, F_{\nu,r}, r_{c,a}, r_{c,a}, \phi_{c,a}, \phi_{w,a}, F_{\nu,a}, r_{c,p}, r_{w,p}, \phi_{c,p}, \\ F_{\nu,p}, r_{w,G}, F_{\nu,G}, x_0, y_0, i, p.a.\}$. 

We determine which combinations of model components to include for each source by considering the features seen in the images, azimuthally averaged visibility profiles, and residuals of less complex models, and adding components to match those features, but also use a quantitative metric (see Table \ref{table:analytic_best_fits}) for each model to help assess the relative likelihood of each model considered. We follow a Bayesian approach and use the multi-nested sampler \texttt{dynesty} \citep{Speagle2019a} to sample a large range of parameter space to determine best-fitting model and derive the shape of the posterior. Samples are drawn from uniform priors on all parameters, with limits listed in Table \ref{table:analytic_priors}.

\subsection{Analytic Modeling Results}
\label{section:analytic_results}

\input{table3.tex}

We show the best fit analytic models compared with our ALMA data in Figures \ref{fig:analytic_fits1} \& \ref{fig:analytic_fits2}, and list the best-fit model parameters in Table \ref{table:analytic_best_fits}. We use the model with the maximum posterior probability from the samples generated by \texttt{dynesty} as the best-fit parameter values, while the listed uncertainties represent the difference between the best-fit parameter values and the 95\% inclusion interval from the samples. For each source we only list the best-fit parameter values for the most-complete model in Table \ref{table:analytic_best_fits}, and we only show model/residual images for that same model in Figures \ref{fig:analytic_fits1} \& \ref{fig:analytic_fits2}. We do, however, show the visibility profiles for all models considered for each source, and we provide a calculation of Bayes Factor in Table \ref{table:analytic_best_fits} for each model considered, compared with the most complete model, for a quantitative assessment of the relative quality of fit for each model.

\begin{figure}
    \centering
    \includegraphics[width=3.25in]{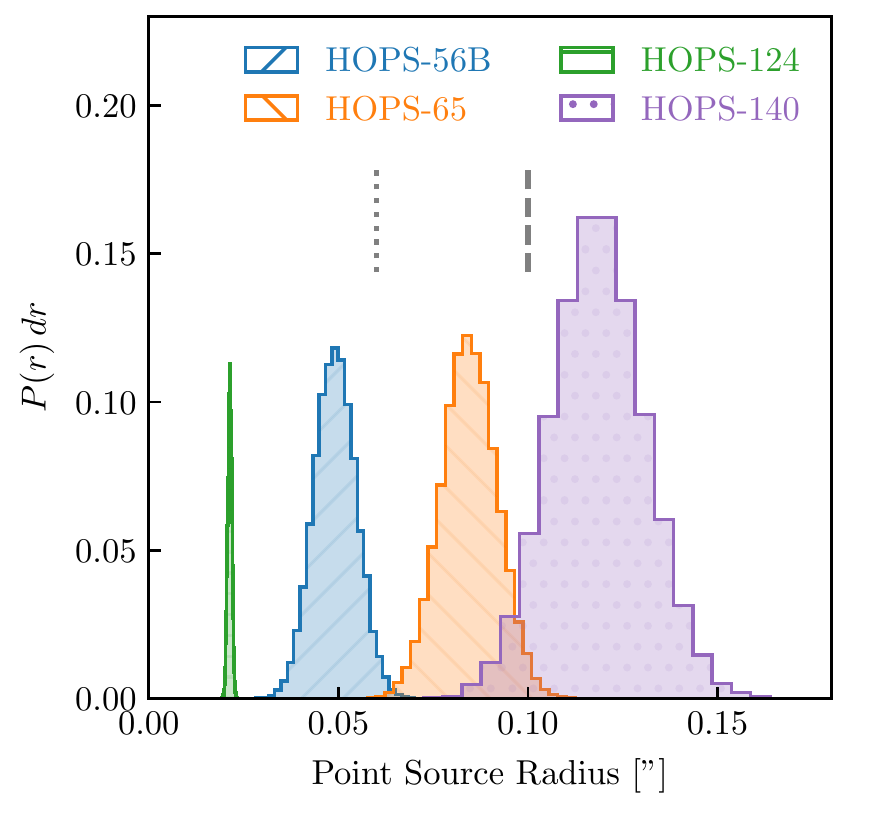}
    \caption{The posterior probability density function for the central point source radius generated by our \texttt{dynesty} fits to four sources that show evidence of central point-like features. In all cases our fits suggest that the point source component is not located at the centroid of the ring structure with $>99\%$ confidence. The gray lines show the approximate beam size for HOPS-56B, HOPS-65, HOPS-140 ($dashed$), and HOPS-124 ($dotted$).}
    \label{fig:point_radius_posterior}
\end{figure}

In general we find that our analytic model does a good job of reproducing the structures we see in the ALMA images of our sample. Only one source, HOPS-163, is fit well by our base ring model alone. HH270-MMS2 is best fit by the base ring model with a large scale Gaussian component, but the remaining five sources require multiple additional layers to adequately reproduce the observed structures. Interestingly, the asymmetric ring with a point source component model for HOPS-65 leaves a significant residual on the northern side of the disk, possibly suggesting that an additional asymmetry is needed to reproduce the disk. The radii of the centers of the rings range from $\sim0.1"-0.6"$ ($40-240$ au at the distance of Orion), with the size of the smallest cavities detected likely limited by the resolution of our observations ($\sim0.1"$).

Four of the sources (HH270-MMS2, HOPS-124, HOPS-140, and HOPS-157) require a Gaussian component on scales larger than the size of the rings component to fit their visibilities at short baselines, which the ring+asymmetry models cannot reproduce. This is of particular note because this large scale emission, that is not evident in the images because it is too low surface brightness, might be associated with the envelopes of these young sources. The full width at half maximum sizes of this component for those four sources ranges from $0.66"-2.6"$, or $\sim250-1000$ au at the distance of Orion. The larger end of that range is much larger than the sizes expected of disks, making the association with an envelope quite likely. We estimate the amount of material that is in these large scale Gaussian components using the standard assumptions of optically thin, isothermal (T $=20$ K) dust with an opacity of 1.84 cm$^2$ g$^{-1}$ \citep[from][at 0.87 mm]{Ossenkopf1994,Tobin2020}, and a gas-to-dust ratio of 100 to convert the millimeter flux to a mass \citep[e.g.][]{Hildebrand1983}. This simple method estimates a range of 0.014 -- 0.32 M$_{\odot}$ of material in these components of dust+gas on large scales.

Four sources (HOPS-56B, HOPS-65, HOPS-124, and HOPS-140) show evidence for central point sources interior to the dominant ring structure. More interestingly, a by-eye inspection of several of the images suggests that this central point source is not located at the center of the ring structures. To explore this possibility for sources that included a central point source in their analytic models, we show the posterior probability density function (PDF) generated by \texttt{dynesty} in Figure \ref{fig:point_radius_posterior}. For all four of these sources, the posterior PDF suggests, with high degrees of confidence, that the central point source component of the fit is offset from the center of the ring component; none of the PDFs are consistent with the point source component being centered at the same location. We do, however, note that for HOPS-140, the inner point source falls very close to the inner edge of the ring. Given the low signal-to-noise ratio of the image, it is possible that this feature is simply a part of the disk that our eyes are drawn to because of the noisiness of the image.

\subsection{Radiative Transfer Modeling}
\label{section:rt_models}

\begin{figure*}[t]
    \centering
    \includegraphics[width=7in]{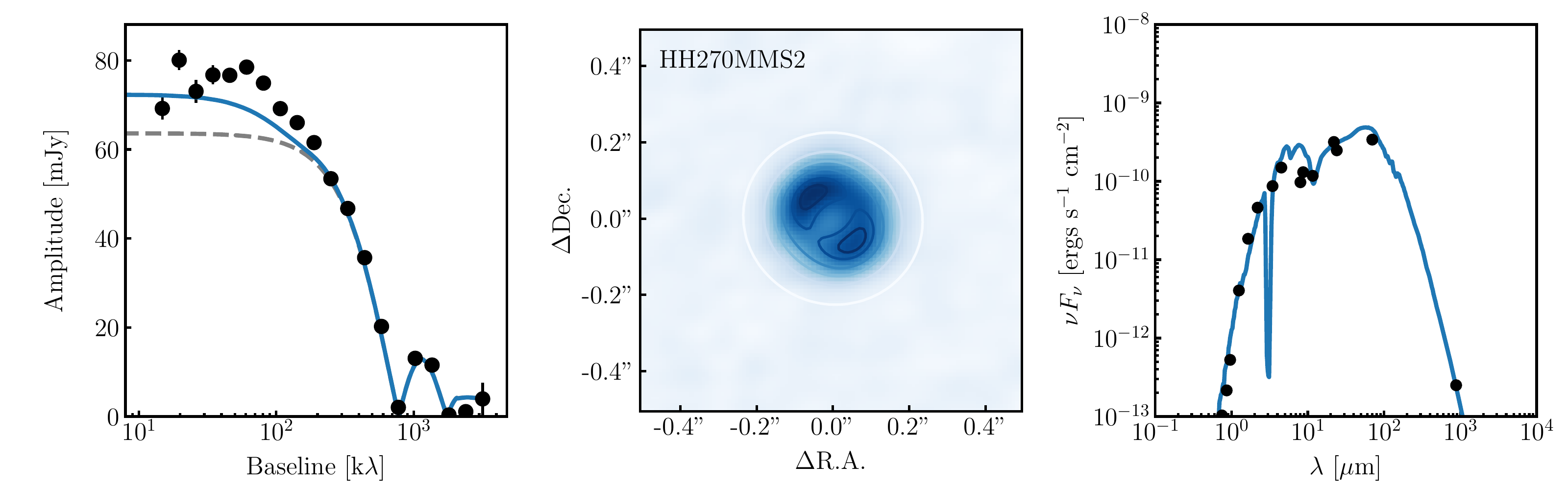}
    \includegraphics[width=7in]{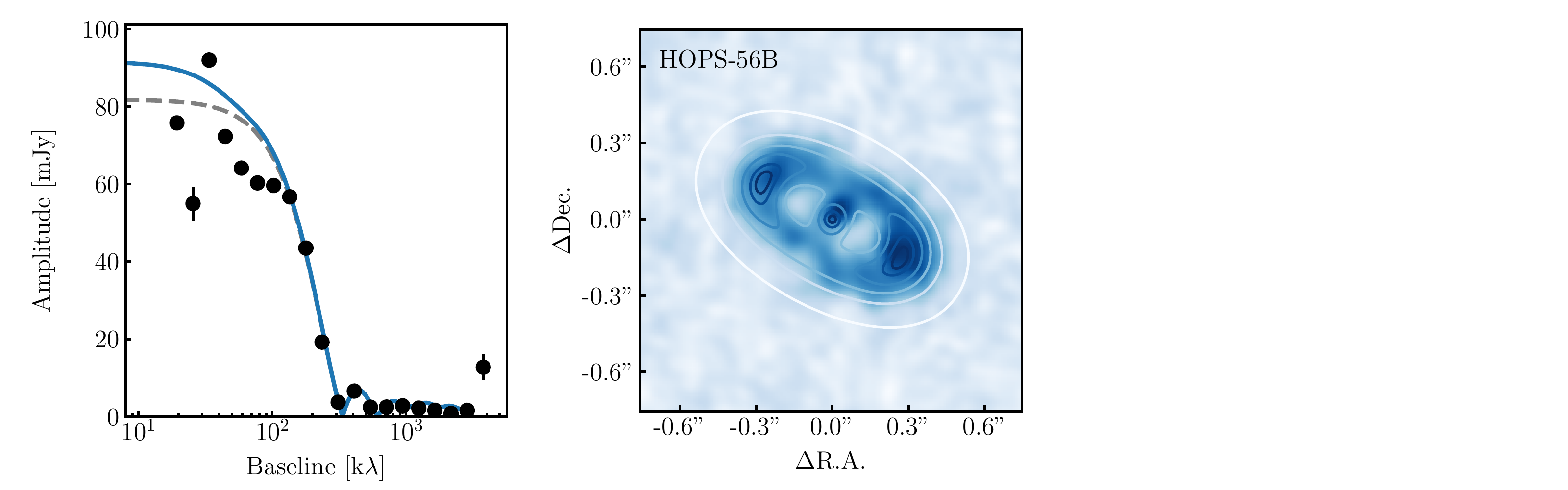}
    \caption{Examples of our radiative transfer model modeling results. The leftmost column shows the azimuthally averaged visibilities with the best-fit radiative transfer modeling in blue. The grey dashed line shows the disk contribution to the best-fit model. The center column shows the ALMA 870 $\mu$m image with the best fit model in contours. The contours show emission at 5\% (white), 25, 45, 65, 85, and 95\% (darkest blue) of the peak value in the model image, and are meant to show that the model does a good job of reproducing the features seen in the data. And the right column shows the broadband SED (and IRS spectrum when available) along with the best-fit model again in blue. Additional plots of the data, model, and residual images are available in Section \ref{section:appendix} of the Appendix.}
    \label{fig:rt_fits1}
\end{figure*}

In addition to our simple analytic models, we fit protostar+disk+envelope radiative transfer models to these targets to try to better characterize their structure in a physically motivated way. Our modeling follows the methods described in \citet{Sheehan2017b}, with a few minor updates and differences. In short our model includes a protostar with a temperature of 4000 K, which is reasonable for a low-mass protostar ($\sim0.5$ M$_{\odot}$; though the protostar masses are not known a-priori), and a luminosity that is left as a free parameter, a power-law surface density disk with an exponential cutoff motivated by viscous disk evolution \citep[][modified from a power-law surface density disk truncated at $R_{disk}$ used in \citealt{Sheehan2017b}]{LyndenBell1974}, and a rotating collapsing envelope described by \citet{Ulrich1976}. 

Unlike \citet{Sheehan2017b}, we fix the inner radius of the disk at 0.1 au, and include a central cavity or a gap with some amount of depletion. The choice of whether to include a gap or a cavity in the radiative transfer model was driven primarily by whether a central point source was visible in the ALMA images, but also informed by the results of the analytic model fitting as to which components were needed to fit the data. A cavity adds two parameters: the outer radius of the cavity ($R_{cav}$) and the factor by which the density inside the cavity is multiplied ($\delta_{cav}$). Similarly, a gap adds three parameters: the radius of the gap center ($R_{gap}$), the width of the gap ($w_{gap}$), and the factor by which the density is reduced inside the gap ($\delta_{gap}$). Finally, in the absence of measured distances for individual sources, we assume they are all at 400 pc, a typical distance for Orion \citep{Zucker2019}.

\begin{figure*}
    \centering
    \includegraphics[width=7in]{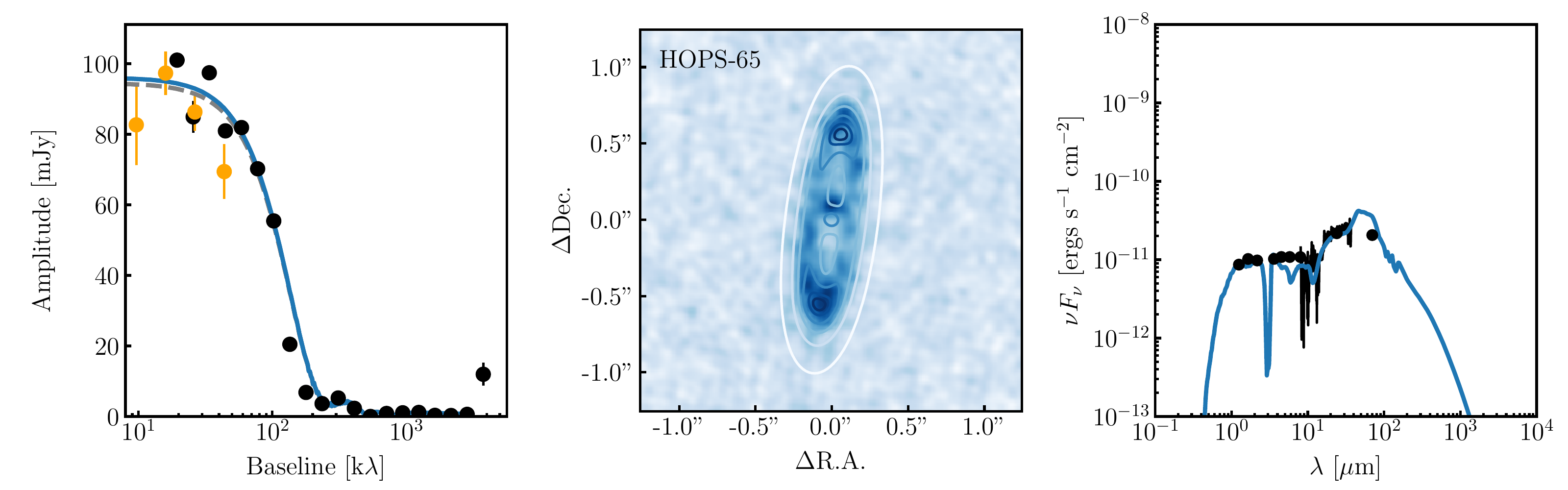}
    \includegraphics[width=7in]{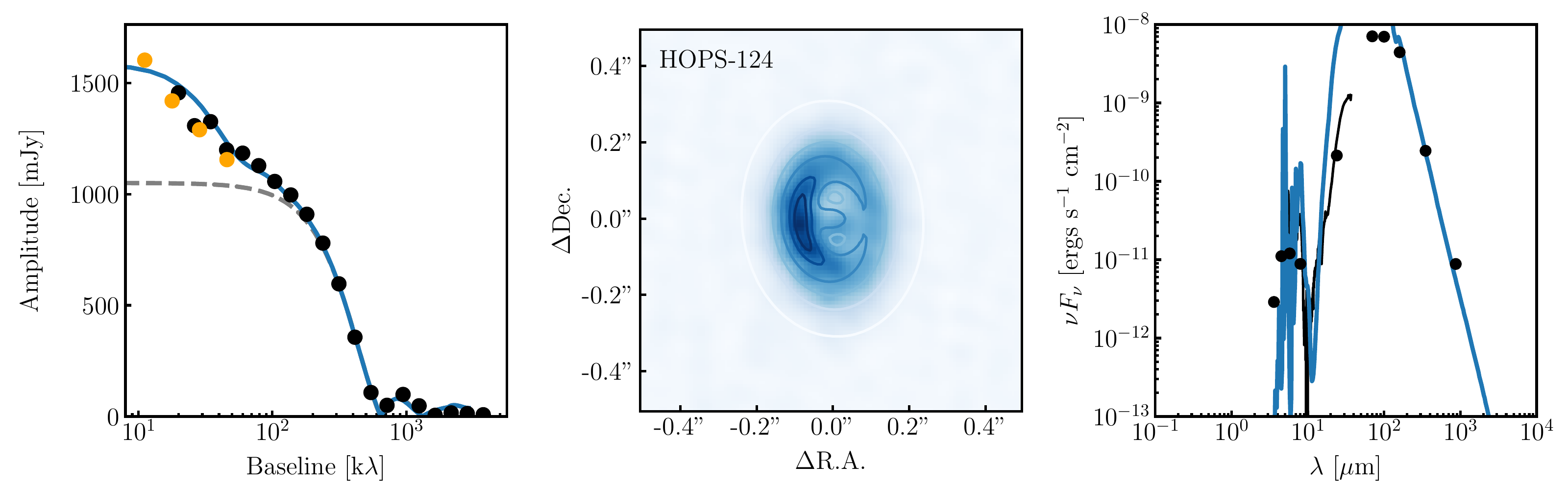}
    \caption{The same as shown in Figure \ref{fig:rt_fits1} for two additional sources. Additionally, the yellow points show the one-dimensional, azimuthally averaged ACA visibilities. Though these data were not included in the fits, they are shown here to demonstrate that our models are consistent with observations extending to even larger scales than probed by our ALMA main array data.}
    \label{fig:rt_fits2}
\end{figure*}

We note that these radiative transfer models assume that the disk is axisymmetric - we make no attempts to include non-axisymmetric structure in the models. Though it would be possible to specify an appropriate model density distribution, the increase in computational cost to go from a two dimensional, axisymmetric radiative transfer model to a full three dimensional radiative transfer model is prohibitive. As such, here we do not seek to reproduce the non-axisymmetric features such as the disk asymmetries, or the central point source components that may be offset from the center of the ring. Nonetheless, these models provide a way to estimate disk and envelope properties in a more physically motivated way.

We use the RADMC-3D code \citep{Dullemond2012} to run the radiative transfer and generate a temperature structure for the prescribed density distribution, and then to subsequently generate synthetic observations, including broadband spectral energy distributions (SEDs) and 345 GHz millimeter images. We use the \texttt{GALARIO} code \citep{Tazzari2018} to Fourier transform the millimeter image into the visibility plane to compare directly with our data. Finally, as we did for the analytic modeling, we follow a Bayesian approach to sampling parameter space, though here we instead use the \texttt{emcee} code to run a Markov Chain Monte Carlo (MCMC) fit of the synthetic observations to our dataset. Full details of the modeling can be found in \citet{Sheehan2017b}.

For most sources, we fit these models to our combined SED + ALMA main array visibilities dataset, not including the additional ACA observations, which were unavailable at the time. One source, HOPS-56B, however, is a multiple with a close enough separation between the companions that it is difficult to disentangle the photometry of each individual component, particularly for the $Herschel$ data at longer wavelengths. Because of this, we only fit the millimeter visibilities of HOPS-56B and do not include any additional information. Furthermore, prior to fitting models to our data, we center the visibilities for each source using the centroid found from the best-fit analytic models. Though we explored leaving the centroid as a free parameter, we found that the inability of our models to reproduce non-axisymmetric structure often affected the resulting best-fit model, and decided that the analytic models do a better job of finding the appropriate center.

\begin{figure*}
    \centering
    \includegraphics[width=7in]{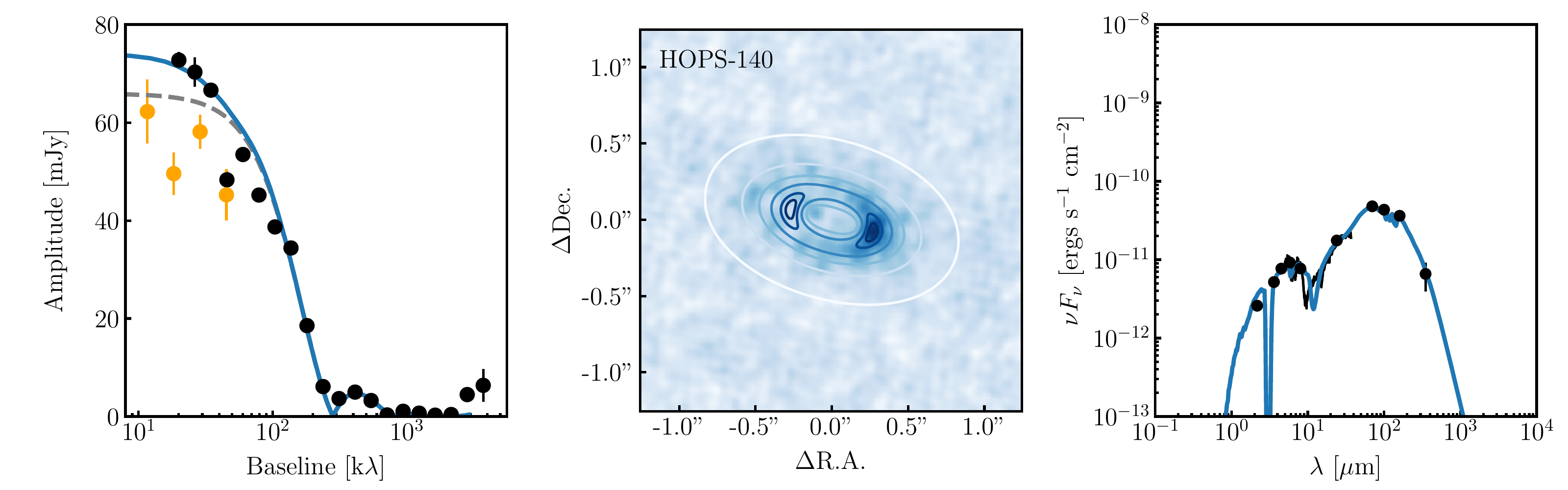}
    \includegraphics[width=7in]{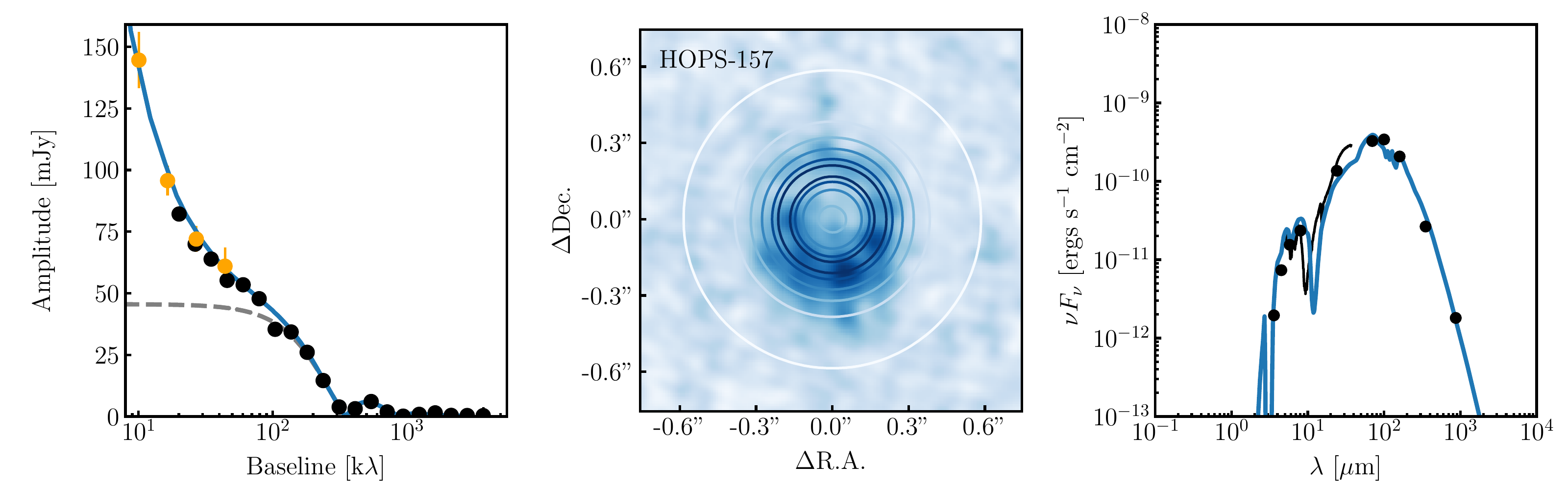}
    \includegraphics[width=7in]{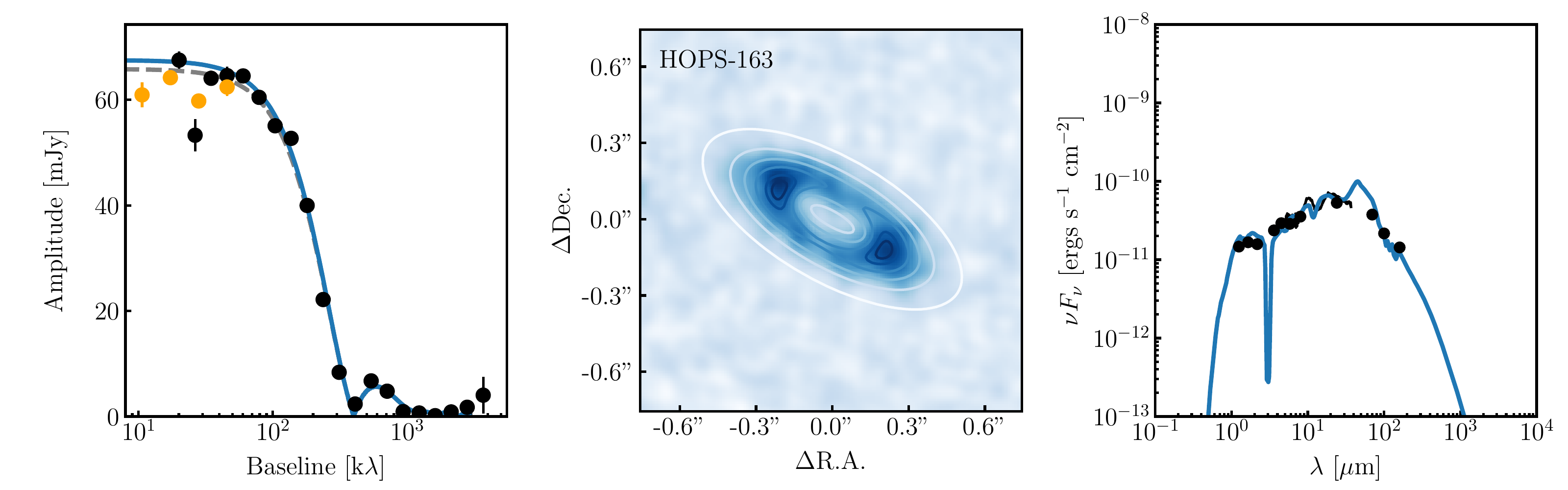}
    \caption{The same as shown in Figure \ref{fig:rt_fits1} for three additional sources.}
    \label{fig:rt_fits3}
\end{figure*}

\subsection{Radiative Transfer Modeling Results}
\label{section:rt_results}

We show the best fit radiative transfer models compared with the data in Figures \ref{fig:rt_fits1}, \ref{fig:rt_fits2}, and \ref{fig:rt_fits3}, and list the best fit parameter values and uncertainties in Table \ref{table:rt_best_fits}. To save space we only show the models as contours here, but we also include additional plots showing data, model, and residual images for the radiative transfer models in Section \ref{section:appendix} in the Appendix. In the panels showing the azimuthally averaged visibility data, we include the short baseline data from the ACA separately, shown in yellow, for sources that had such data available. Despite not being included in the fits, our best fit models show good consistency with the short-baseline ACA data, which gives us confidence that we are recovering large scale structure well with our models.

\input{table4.tex}

To calculate the best fit parameters and uncertainties, we collect every step of every MCMC walker post-convergence to have a collection of samples. Walkers that got ``lost" \citep[likely in local minima, see the Appendix of][for further details]{Sheehan2019}, were redistributed around the median of the post-burn-in samples, and additional steps were run to allow them to converge, ensuring that we are properly sampling the peak of the posterior probability distribution. The best fit parameters are determined by the model with the maximum posterior probability, and the uncertainties are calculated from the difference between the best-fit model parameters and the range including 95\% of the samples.

Interestingly, despite assuming an axisymmetric structure, our model is able to reproduce the asymmetry in HOPS-124's disk quite well. We have explored adjusting the parameters of this model to determine the underlying cause of this asymmetry in an otherwise symmetric model, and it appears to be the result of of massive, flared disk viewed at an appropriate inclination angle. The high mass of the disk means that we are likely seeing material closer to the surface of the disk, and because of the large scale height and inclined viewing angle, on the far side of the disk we see the surface layer directly, while the near side we see extincted through colder outer regions of the disk, creating an asymmetry in the emission. Such line-of-sight effects should appear along the minor axis of the disk, as is the case with HOPS-124. The remaining sources have asymmetries that are either along the major axis, or are at an intermediate position angle, and so it is not surprising that our model fails to reproduce them.

\begin{figure*}
    \centering
    \includegraphics[width=7in]{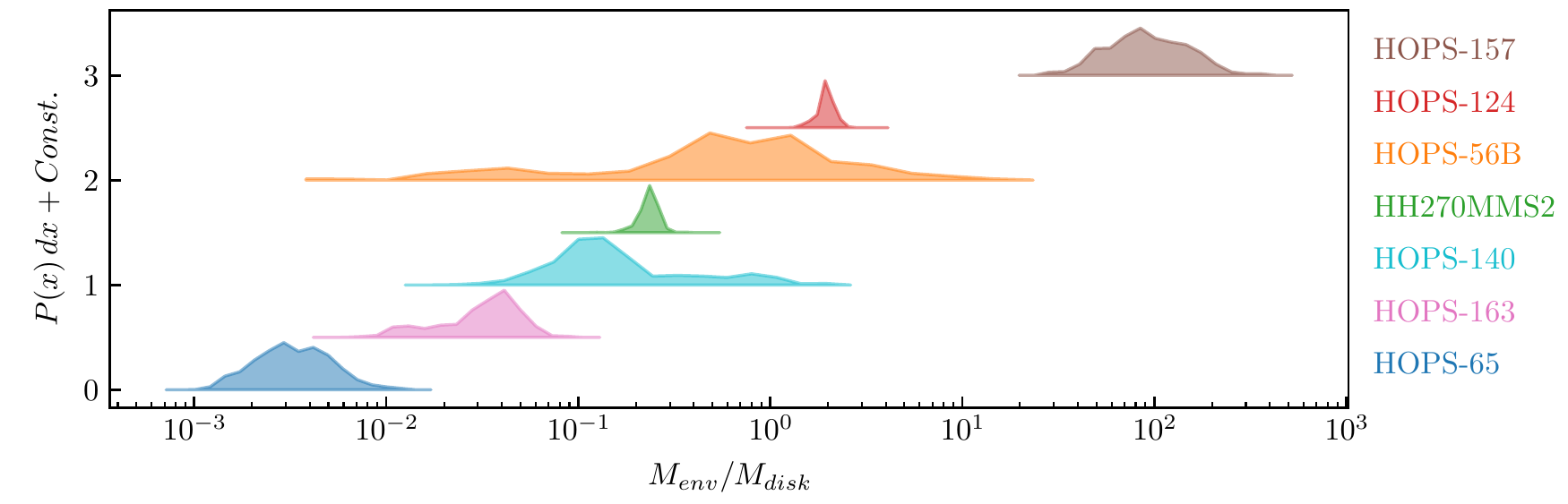}
    \caption{Posterior distributions for $M_{env}/M_{disk}$ found from our radiative transfer modeling. Each distribution is offset from zero by some amount for ease of viewing. If we assume that $M_{env}/M_{disk}$ represents an evolutionary sequence, then larger values should represent younger sources.}
    \label{fig:MenvMdisk_posterior}
\end{figure*}

In Figures \ref{fig:rt_fits1}, \ref{fig:rt_fits2}, and \ref{fig:rt_fits3} we show not just the best fit model curve (blue) but also the contribution of the disk component to the visibilities as a grey dashed line. For the same four sources that needed large scale Gaussian components in the analytic modeling (HH270-MMS2, HOPS-124, HOPS-140, and HOPS-157) we see quite large excesses of emission at short baselines (large spatial scales) compared with the disk contribution to the the model. This excess emission comes from large scale envelope emission, and indicates that these sources are embedded within particularly massive envelopes as compared with their disks.

Our radiative transfer model fit to HOPS-56B also shows evidence that the disk is embedded within a massive envelope. We do, however, note that HOPS-56B is nearby to a deeply embedded source for which the extended envelope can even be seen in the ALMA image. As such, it is possible that the envelope we measure here results from an incomplete subtraction of the extended emission from the nearby source, and very well could over-estimate the mass of HOPS-56B's envelope. We therefore caution that the results for HOPS-56B should be considered more uncertain than the nominal errors show.

The typical young stellar object classification scheme (Class 0, I, II, III) is thought to represent an evolutionary sequence, with the youngest and most significantly embedded disks classified as Class 0, and the oldest with no envelope or disk remaining as Class III \citep[e.g.][]{Chen1995,Lada1987,Myers1987}. Most of these sources are classified as Class I embedded protostars, while HOPS-124 is classified as Class 0, based on these standard classification methods such as near-infrared spectral index \citep[e.g.][]{Furlan2016}. However, such methods can have uncertainty in the underlying physical structure; for example, an edge-on protoplanetary disk might look like an embedded disk \citep[e.g.][]{Chiang1999,Crapsi2008}. Moreover, they provide a relatively limited picture of the evolutionary state of a source.

Our radiative transfer modeling, however, provides a means to make a more physically motivated classification of our sources. As envelope mass is expected to decrease with time, we suggest using the ratio of the envelope mass to the disk mass ($M_{env}/M_{disk}$) as an evolutionary indicator \citep[e.g.][]{Robitaille2006,Crapsi2008}. In this scheme, sources with $M_{env}/M_{disk} \gg 1$ are embedded in massive envelopes and very young. Sources with $M_{env}/M_{disk} \sim 1$ are embedded but much less substantially, and sources with $M_{env}/M_{disk} \ll 1$ are likely at the end of the embedded phase, or are even already protoplanetary disks. Both classification schemes seek to quantify the same thing: how substantial the protostellar envelope of a source is. However with the additional information provided by our modeling our proposed scheme relies on more direct measurements of physical properties.

It's important to note that while this scheme provides a physically motivated way to estimate the relative age of protostars, care should be taken in over-interpreting the values as a direct measurement of age. It is unlikely that the mapping from $M_{env}/M_{disk}$ to age is linear, or even necessarily monotonic, as the rate at which material from the envelope accretes onto the disk and the rate at which material from the disk accretes onto the star can vary substantially \citep[e.g.][]{Vorobyov2010,Bate2018}. Additionally, \citet{Tobin2020} showed that $M_{disk}$ does not have a strong systematic dependence on evolutionary state, so it is similarly difficult to know exactly how $M_{env}/M_{disk}$ traces evolution. Still, broadly speaking, sources with large values of $M_{env}/M_{disk}$ should be at a relatively early evolutionary state and therefore are likely quite young.

To apply this to our sample of protostars, we show the posterior on $M_{env}/M_{disk}$ in Figure \ref{fig:MenvMdisk_posterior}. From our radiative transfer modeling we find that our sources have $M_{env}/M_{disk}$ ranging from $0.003^{+0.005}_{-0.002}$ (HOPS-65) -- $94.38^{+121.86}_{-56.13}$ (HOPS-157). Following our proposed evolutionary indicator scheme, we suggest that HOPS-65 is likely a very late stage embedded source, close to emerging from it's envelope, if indeed there's any envelope left at all. Conversely, HOPS-157 appears to be very significantly embedded, and therefore quite young. HOPS-124 has $M_{env}/M_{disk} = 1.95^{+0.40}_{-0.44}$, and is likely quite young and embedded as well, while HH270-MMS2 has a moderate value, of $M_{env}/M_{disk} \sim 0.23^{+0.06}_{-0.06}$. HOPS-140 has a lower value, of $0.13^{+0.96}_{-0.08}$, suggesting it may be disk-dominated, but with large errors that suggest it could potentially be more moderately embedded, with $M_{env}/M_{disk} > 1$. HOPS-163 has $M_{env}/M_{disk} = 0.037^{+0.024}_{-0.027}$, suggesting that while there may be some envelope left, it is dominated by its disk and will likely soon emerge from its envelope as a protoplanetary disk. Lastly, our modeling suggests that HOPS-56B has a moderate value for $M_{env}/M_{disk}$, but as was previously mentioned the envelope mass reported by this modeling should be considered particularly uncertain, and so the so the evolutionary state of this source is unclear.

Finally, we note that for HOPS-124 and HOPS-157, the envelope masses measured here, and consequently their $M_{env}/M_{disk}$ ratios, might reasonably be considered lower limits. Our ALMA observations, if the ACA data are included, are sensitive to spatial scales of $\sim8000$ AU. Our models reproduce the total flux on those scales quite well, including the ACA observations despite their not being included in the fit, suggesting that we are recovering all of the mass out to $\sim8000$ AU scales. There may, however, be additional emission on larger scales that is not accounted for by our observations. Indeed, at the shortest baselines in our ALMA data, the visibilities for HOPS-124 and HOPS-157 are still rising, indicating that there is more material on still larger scales than those probed by our observations. Our models do, to some degree. account for this, as the model visibility profiles are also rising at the shortest baselines. However, our HOPS-124 model underpredicts the APEX/LABOCA measured 870 $\mu$m flux, suggesting that there could be additional material that we are not accounting for. Our model for HOPS-157 also underpredicts the single dish flux, though less severely, so we may be more severely underestimating the $M_{env}/M_{disk}$ ratio for HOPS-124.

Though the same could be true for the remaining sources, their envelope masses are already quite low and so the likelihood that there is enough material on larger scales to substantially alter our measured $M_{env}/M_{disk}$ seems small. The APEX observations for these remaining sources, when available, were all upper limits, and their visibility profiles at the shortest baselines are flat, further suggesting that we aren't missing a significant amount of material.

\section{Discussion}
\label{section:discussion}

Our modeling suggests that many of the sources in our sample are quite young, with $M_{env}/M_{disk} \gtrsim 0.5$. For comparison, GY 91, one of the first embedded sources found to have substructures, has $M_{env}/M_{disk} \sim 0.1$ \citep{Sheehan2018}. \citet{Sheehan2017a} also suggested that WL 17, a young embedded source with a large cavity or hole, is similarly a late-stage embedded source. As such, the sources in this sample represent the youngest known disks to-date that show substructures.

\begin{figure*}
    \centering
    \includegraphics[width=7in]{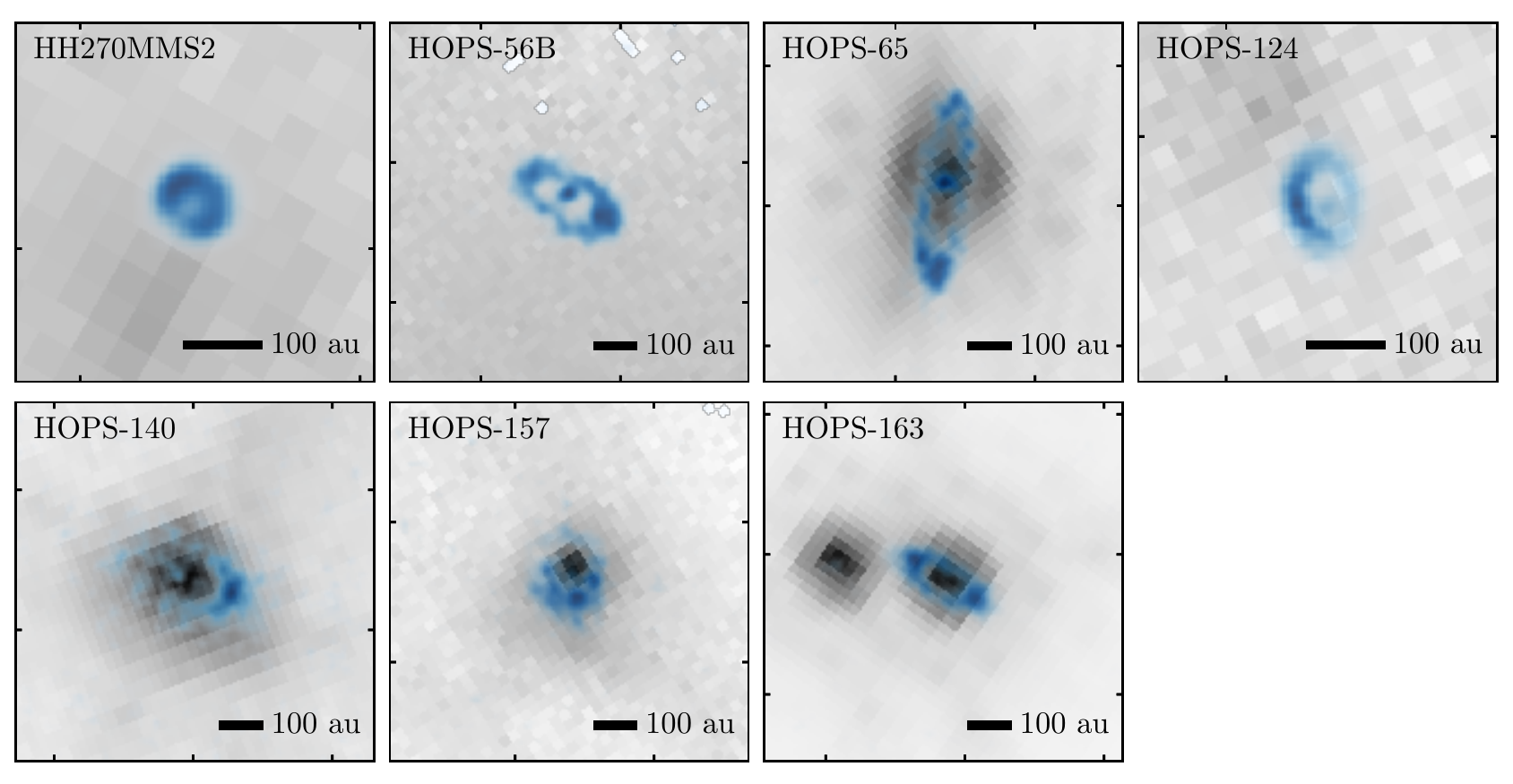}
    \caption{HST NICMOS or WFC3 near-infrared imaging of the seven protostars in our sample ({\it greyscale}) with our ALMA images overlaid ({\it blue}). A line corresponding to 100 au is shown in each image, for scale. Though a few sources appear to be multiples on larger scales, there is no evidence for multiple stars falling within the cavities of the disks.}
    \label{fig:alma_hst}
\end{figure*}

What is unclear, however, is how substructures can form at such early times. Photoevaporation typically takes time before it can significantly alter the structure of the disk \citep[e.g.][]{Alexander2006,Gorti2009a,Owen2010,Armitage2011}, and as these sources are young and have high accretion rates, it seems unlikely that photoevaporation could create these large holes. Dust grain growth has been proposed as another possibility \citep[e.g.][]{Dullemond2005}, though in practice it is difficult to make cavities that show up in millimeter images \citep[e.g.][]{Birnstiel2012}. Other explanations such as variations in dust properties near snowlines \citep[e.g.][]{Clarke2001,Zhang2015,Okuzumi2016} or magnetic zonal flows \citep{Flock2015} are more difficult to test or disprove, though they have not been shown to produce large-scale asymmetries either. Snowlines do not appear to be responsible for the bulk of the substructures found in protoplanetary disks \citep[e.g.][]{Long2018,vanderMarel2019,Huang2018}, but could explain some features. Still, this does not mean that they can be ruled out for younger disks.

Disk winds driven by magnetorotational instability turbulence have been shown to be capable of clearing central cavities in young sources \citep{Takahashi2018}. Such winds could therefore explain sources with central cavities, like HH270-MMS2 and HOPS-163. It is not clear, however, whether such models could explain the azimuthal asymmetry seen in HOPS-157. Moreover, the models of disk winds do not leave an inner disk, and so they may not be able to reproduce the structures seen in the remaining sources in our sample.

One of the most appealing possibilities is the presence of unseen massive bodies shaping the disk dynamically. In the case of protoplanetary disks, these massive unseen bodies are often thought to be planets \citep[e.g.][]{Dong2015,Isella2016}. It is tempting to associate the features found in our protostellar disks with planets, particularly as planet formation is thought to start early - protoplanetary disks do not typically have enough mass to form the observed planet distribution \citep[e.g.][]{Najita2014,Manara2018}, whereas protostellar disks might \citep[e.g.][]{Sheehan2017b,Andersen2019,Tobin2020}. Many of the gaps and cavities found in our protostellar disk sample, however, are quite large, with widths as large as $\sim200$ au. Moreover, some of our sources appear to be quite young, and though planet formation is thought to start early, it is not clear how early on planets can form \citep[e.g.][]{Drazkowska2018}.

Gravitational instabilities in disks could potentially form planets on significantly shorter timescales than core accretion \citep[e.g.][]{Boss1997,Durisen2007,Boss2011}, and could therefore explain how these features are seen at such early times. Indeed, HOPS-124 underwent a flare circa $\sim$2009 in the mid-infrared, seen with $Spitzer$ (T. Megeath, private communication, in press). Outbursts seen in other young protostars \citep[e.g.][]{Wachmann1939,Safron2015,Fischer2019}, though much larger in magnitude than what was seen for HOPS-124,, have been linked to gravitational instabilities in disks driving a rapid accretion of material onto the central protostar \citep[e.g.][]{Vorobyov2010}. This outburst may provide a clue that gravitational instabilities could be acting in these systems to form unseen massive bodies on short timescales. However, if gravitational instabilities are indeed acting in these sources, simulations suggest that they are more likely to produce stellar or sub-stellar companions than planetary-mass objects \citep[e.g.][]{Kratter2010,Zhu2012,Forgan2013}.

Therefore, we suggest that at least some of these substructures may be indications that we are observing young binary systems with circumbinary disks. Binaries have been shown to reproduce all of the major features that we see here in their disks, including cavities and large asymmetries \citep[e.g.][]{Price2018,Calcino2019,Poblete2019}, and are perhaps more likely to carve the very large gaps and cavities that we see here. They may also naturally explain the offset of the inner disk from the center of the ring: if the masses of the stars in the binary are similar, the stars may orbit significantly offset from the center of mass of the system, and we could be seeing the circum-stellar disk for that source.

To test whether there is any evidence that these are multiple systems, we compare our ALMA observations with Hubble Space Telescope (HST) archival near-infrared imaging with NICMOS or WFC3 \citep{Kounkel2016}. We perform a crude astrometric alignment of the HST images in a few ways. For HOPS-65 and HOPS-163 we shift the images such that sources in the HST images align with $Gaia$ Data Release 2 \citep{GaiaCollaboration2018} detections. For HOPS-140 we shift the HST images such that two stellar detections in the HST image align with two point source detections in the ALMA image. And for HOPS-157 we shift the lone star in the HST image to roughly match with the ALMA disk detection. In all cases the shifts needed to match ALMA and HST images are $\lesssim0.5''$ in distance. The remaining three sources (HH270-MMS2, HOPS-56B, and HOPS-124) did not have sufficient detections in HST imaging to motivate any adjustments, but based on the four sources that did, we expect that their locations should be accurate to within $\sim0.5''$. We show the HST images in Figure \ref{fig:alma_hst} with the ALMA disk detections shown on top.

We do not find any compelling evidence for multiplicity for any of the sources in our sample. HOPS-56B and HH270-MMS2 do not appear to be associated with any emission in their corresponding HST images, likely due to heavy foreground extinction. HOPS-65, HOPS-140, HOPS-157, and HOPS-163 are all associated with point-source detections in their HST images but do not appear to be close separation multiples. HOPS-140 and HOPS-163 do have second point sources in their fields of view, but they are exterior to the transition disks, and they appear to both be associated with very weak point source ALMA detections. HOPS-124 is not associated with any point sources, but there is extended structure that appears to possibly be associated with an outflow cavity. 

That said, the spatial resolution of the HST images is relatively coarse compared with the sizes of the cavities of most of our sources, and so it is perhaps expected that we would not be able to clearly identify multiplicity for our sample in this manner. Interestingly though, the inner disk of HOPS-65 detected with ALMA is well aligned with the HST point source, both offset from the center of the ring. Moreover, no additional point source is seen within the cavity, despite the cavity being large enough that the HST resolution would not be prohibitive. Still, with the large, extended point spread function of the HST image, any limits are likely to be uninteresting as the contrast between the primary and a putative companion could be too large.

\section{Conclusions}
\label{section:conclusion}

In summary, we find seven protostellar disks (aged $\sim0.1-1$ Myr) with newly detected disk substructures when observed at high angular resolution with ALMA, including central cavities, bright and dark rings, and large scale asymmetries. These disks join a growing population of disks that have been found with substructure, and triples the number of young (Class 0/I) protostars known to have substructure.

To understand the evolutionary stage of these protostars beyond simple, but fallible, evolutionary indicators such as infared spectral index and bolometric temperature, we fit our data for each source with disk+envelope radiative transfer models. We find that the degree of ``embeddedness" varies substantially from source to source, with $M_{env}/M_{disk}$ ranging from $0.003^{+0.005}_{-0.002}$ -- $94.38^{+121.86}_{-56.13}$. A few of the sources have very small values of $M_{env}/M_{disk}$, indicating that they may be late-stage embedded protostars, close to emerging from their envelopes as protoplanetary disks, similar to most of the protostellar disks previously found to have substructures. However several sources have $M_{env}/M_{disk} \sim 1$, and two more have $M_{env}/M_{disk} \gg 1$, indicating that these features can develop while disks are still embedded in a substantial envelope of material and while they are quite young.

We also fit simple analytic models to our data to characterize the detailed geometry of the rings, asymmetries and inner disks. In particular, we find evidence that for the four sources that may have inner disks, their inner disks are offset from the center of the ring.

The presence of substructures so early in the lifetimes of disks raises interesting questions about how substructures are formed. Though dynamical sculpting by planets is a popular (and exciting) option, it is unclear whether planets can form quickly enough to carve out gaps in the youngest of our sources. Given the large widths of some of the gaps/cavities found in our sample (as large as ~200 au), as well as the large disk asymmetries, it seems plausible that many of these disks may be indicators of binary formation at early times. We do not, however, have the data to rule out any of the many methods found so far for generating disk substructures, so further observations will be required to understand the nature of these protostellar disks.

\software{pdspy \citep{Sheehan2018b}, CASA \citep{McMullin2007}, RADMC-3D \citep{Dullemond2012}, emcee \citep{ForemanMackey2013}, matplotlib \citep{Hunter2007}, corner \citep{ForemanMackey2016}, GALARIO \citep{Tazzari2017a}, dynesty \citep{Speagle2019a}}

\acknowledgements We thank the anonymous referee for a careful review and a number of suggestions that helped to improve the manuscript. P.D.S is supported by a National Science Foundation Astronomy \& Astrophysics Postdoctoral Fellowship under Award No. 2001830. L.W.L acknowledges support from NSF Grant No. 1910364. The computing for this project was performed at the OU Supercomputing Center for Education \& Research (OSCER) at the University of Oklahoma (OU). This paper makes use of the following ALMA data: ADS/JAO.ALMA\#2015.1.00041.S, 2018.1.01284.S. ALMA is a partnership of ESO (representing its member states), NSF (USA) and NINS (Japan), together with NRC (Canada), NSC and ASIAA (Taiwan), and KASI (Republic of Korea), in cooperation with the Republic of Chile. The Joint ALMA Observatory is operated by ESO, AUI/NRAO and NAOJ. The National Radio Astronomy Observatory is a facility of the National Science Foundation operated under cooperative agreement by Associated Universities, Inc.

\bibliography{ms.bib}

\appendix

\section{Additional Radiative Transfer Modeling Plots}
\label{section:appendix}

\begin{figure*}[h!]
    \centering
    \includegraphics[width=6.45in]{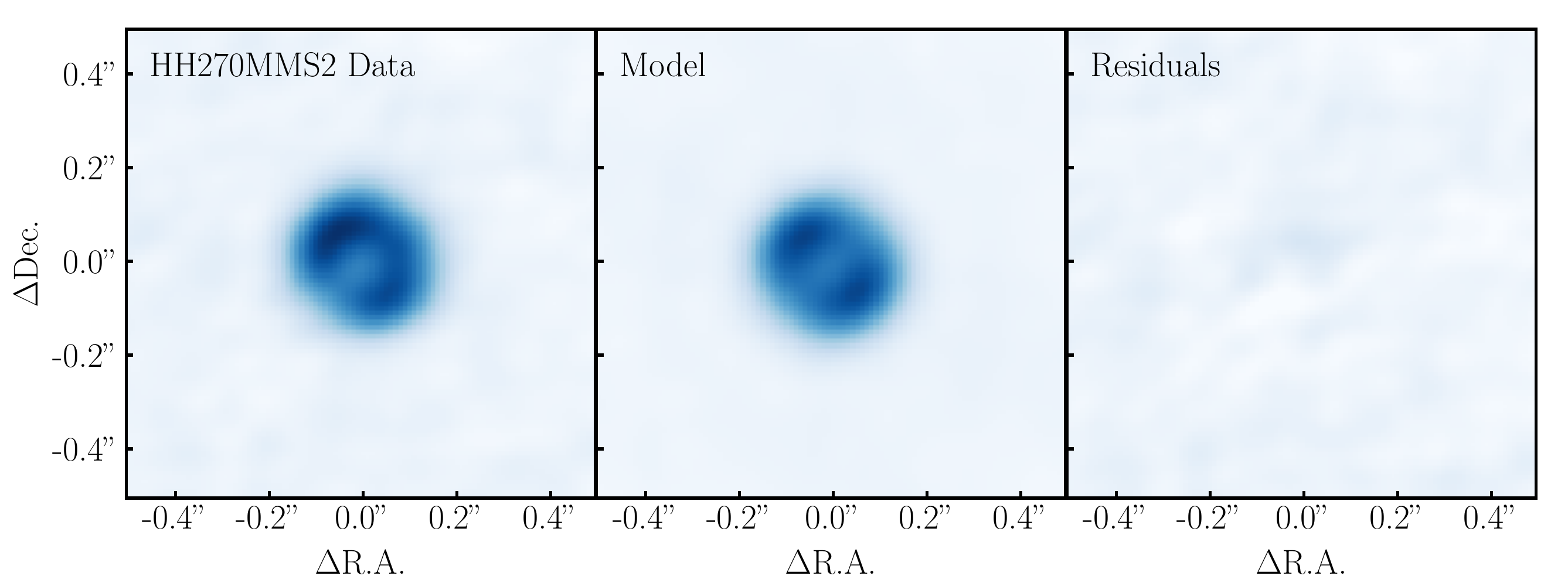}
    \includegraphics[width=6.45in]{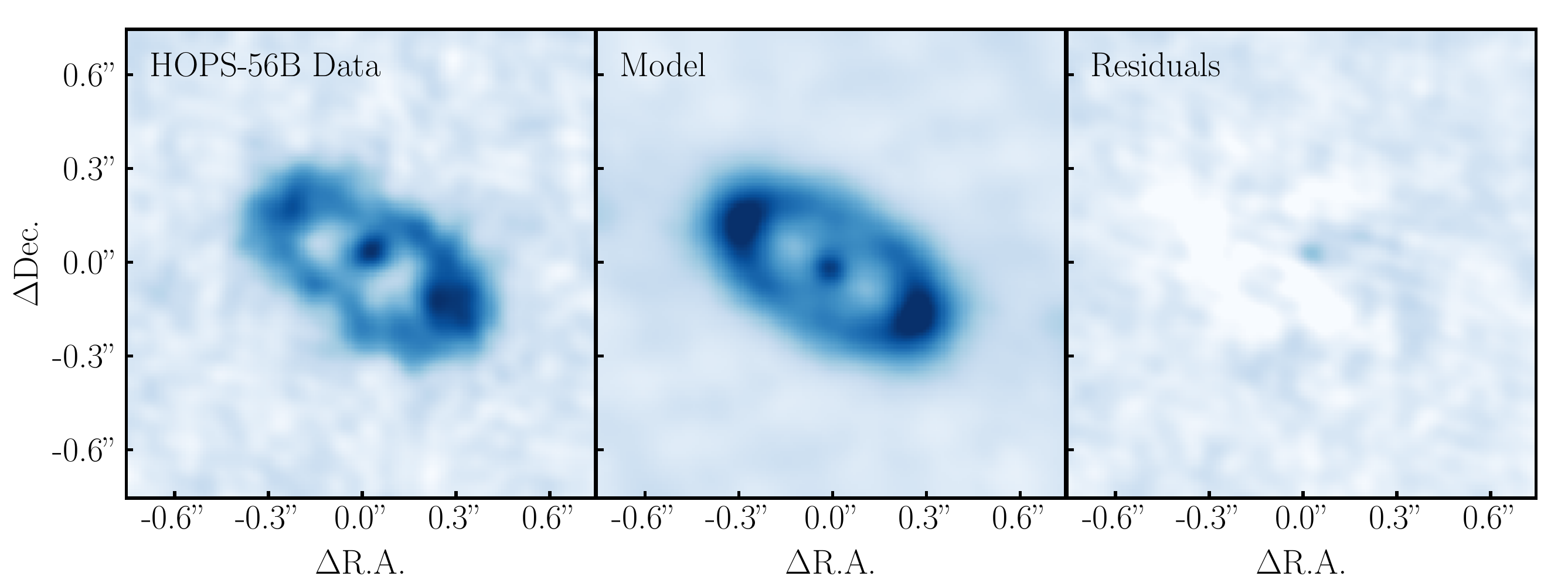}
    \includegraphics[width=6.45in]{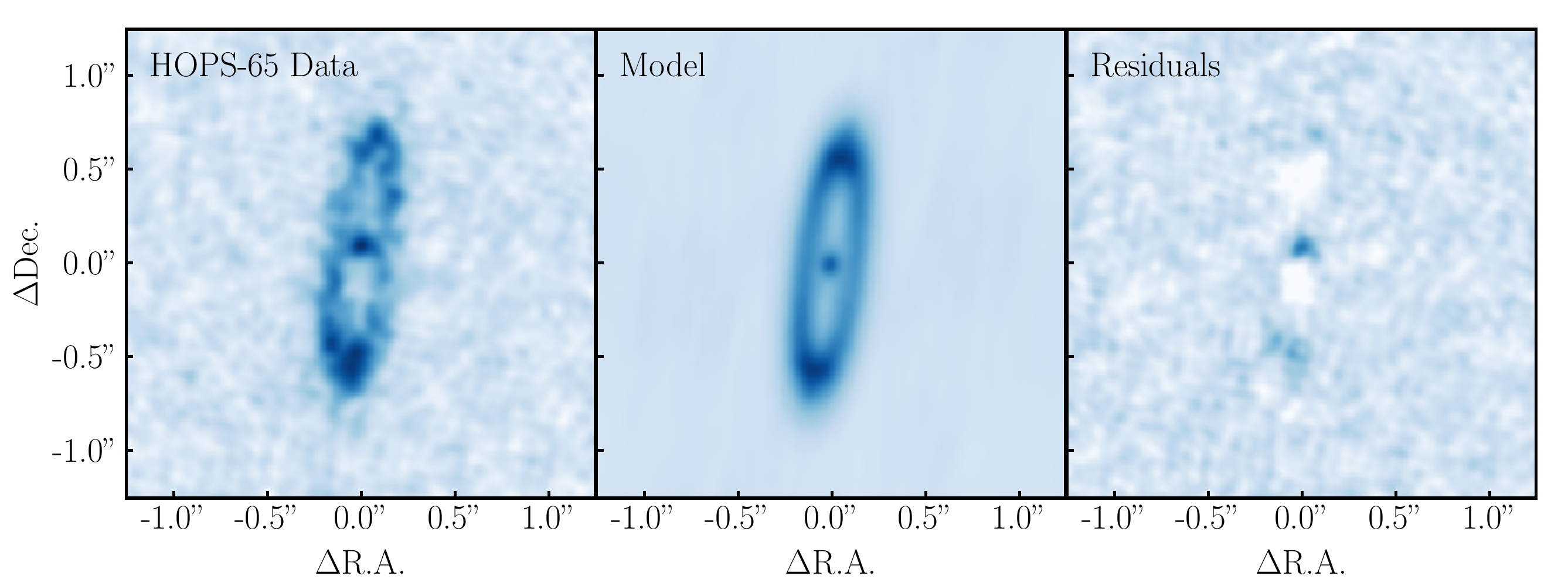}
    \vspace{-5pt}
    \caption{The 345 GHz ALMA continuum images for our sources ($left$) along with a simulated images of the best-fit radiative transfer model ($center$) and the residuals for that model ($right$). The model and residual images are produced by sampling the model at the same baselines as the observations, subtracting the model visibilities from the data for the residuals, and Fourier Transforming those visibilities to produce images. The imaging was done with the same weighting scheme that was used to produce the image of the data, which for most sources was Briggs weighting with a robust parameter of 0.5, while for HOPS-124 we used superuniform weighting.}
    \label{fig:rt_fits_appendix1}
\end{figure*}

\begin{figure*}[h!]
    \centering
    \includegraphics[width=6.45in]{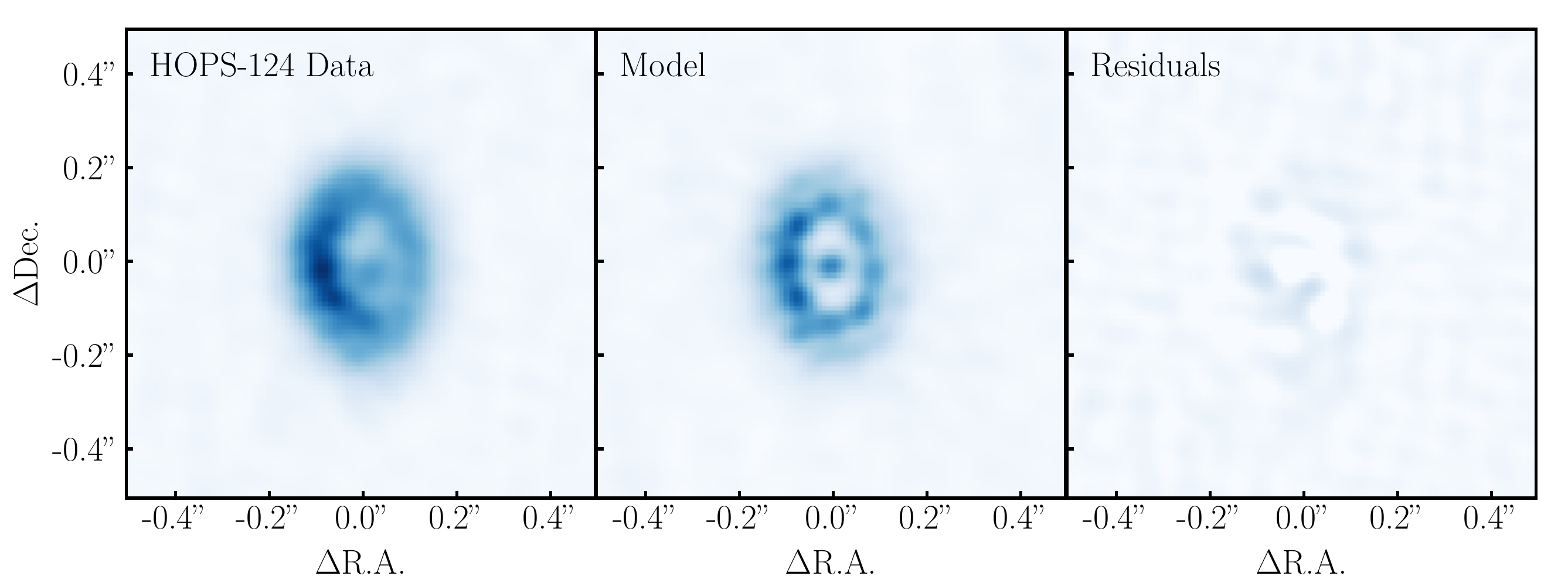}
    \includegraphics[width=6.45in]{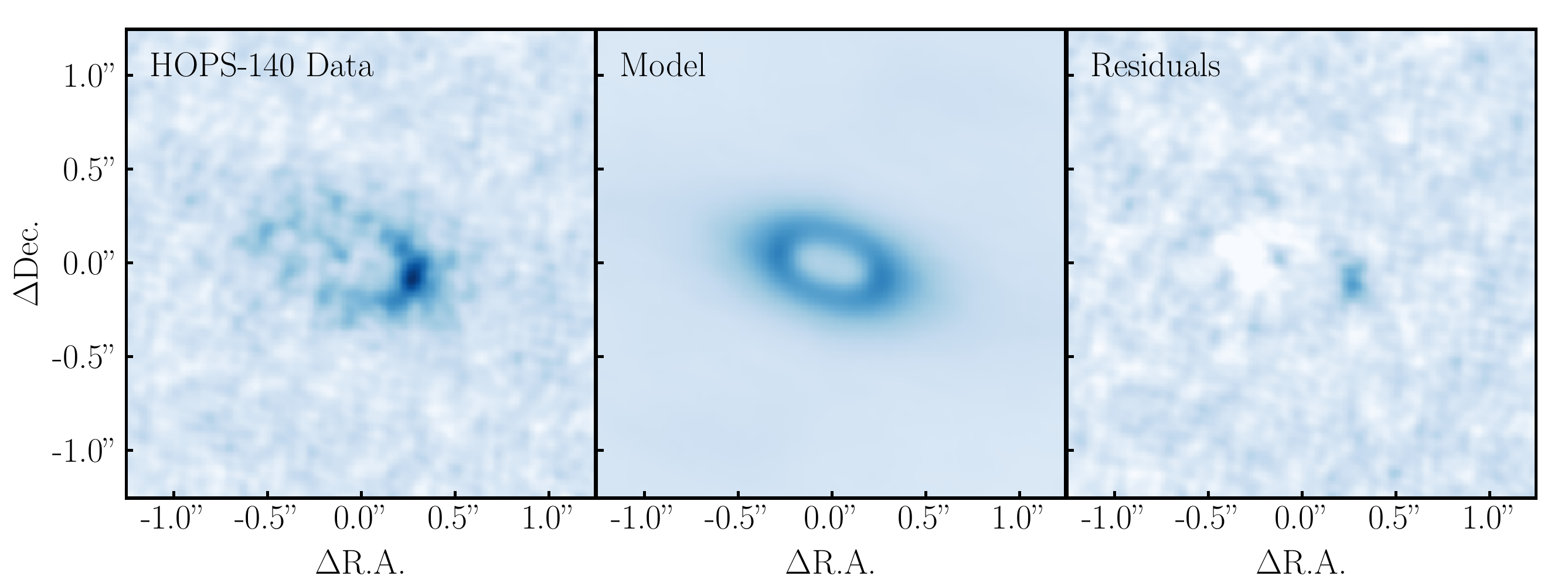}
    \includegraphics[width=6.45in]{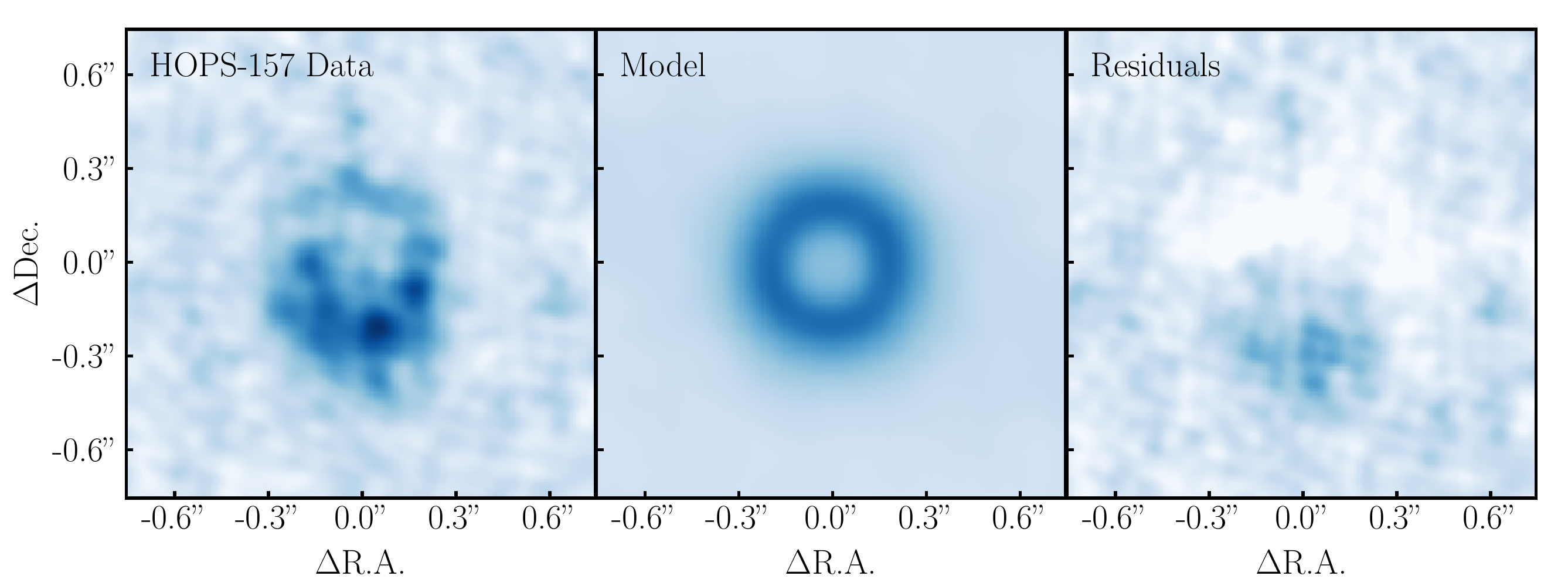}
    \includegraphics[width=6.45in]{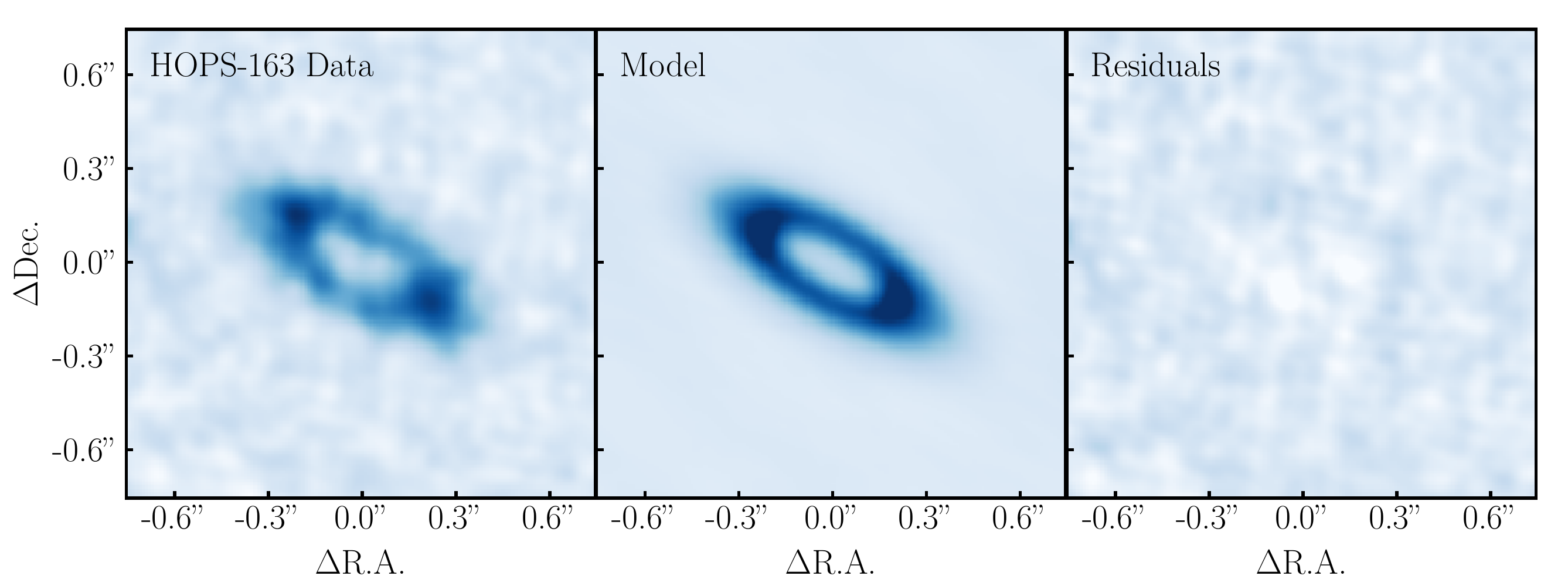}
    \vspace{-5pt}
    \caption{A continuation of Figure \ref{fig:rt_fits_appendix1} for the four remaining sources.}
    \label{fig:rt_fits_appendix2}
\end{figure*}

\end{document}

%% file: table1.tex
\begin{deluxetable*}{lccccccccc}
\tablenum{1}
\tablecaption{Source Properties}
\label{table:source_properties}
\tablehead{\colhead{Object} & \colhead{$L_{bol}$} & \colhead{$T_{bol}$} & \colhead{Class} & \colhead{$F_{0.87{ }mm}$} & \colhead{$F_{9{ }mm}$} & \colhead{Sp. Index} & \colhead{Sp. Index} & \colhead{$F_{0.87{ }mm,ACA}$} & \colhead{$F_{0.87{}mm,APEX}$} \\ \colhead{} & \colhead{[$L_{\odot}$]} & \colhead{[K]} & \colhead{} & \colhead{[mJy]} & \colhead{[mJy]} & \colhead{$0.87$mm - $9$mm} & \colhead{$8.1$mm - $10$mm} & \colhead{[mJy]} & \colhead{[mJy]} }
\startdata
HH270-MMS2  & 4.6 & 249.0 & Flat & $64.2\pm0.7$ & 0.024$\pm$0.024 & $3.39\pm0.43$ & \nodata & \nodata & \nodata \\
HOPS-56B  & 20.4 & 48.1 & 0    & $74.8\pm0.8$ & 0.151$\pm$0.026 & $2.64\pm0.07$ & 1.99$\pm$1.43 & \nodata & \nodata \\
HOPS-65    & 0.3 & 545.7 & I    & $101.6\pm1.3$ & \nodata & \nodata & \nodata & $97.7\pm8.7$ & $<430$ \\
HOPS-124   & 52.3 & 44.8 & 0    & $1086.9\pm0.9$ & 1.450$\pm$0.031 & $2.82\pm0.01$ & 1.01$\pm$0.17 & $1469\pm72$ & $2560\pm510$ \\
HOPS-140 & 0.5 & 137.2 & I    & $60.7\pm1.7$ & \nodata & \nodata & \nodata & $45.1\pm4.8$ & $<340$ \\
HOPS-157   & 3.3 & 77.6 & I    & $49.1\pm1.2$ & \nodata & \nodata & \nodata & $183\pm21$ & $530\pm$104 \\
HOPS-163 & 0.8 & 432.3 & I    & $65.7\pm0.7$ & \nodata & \nodata & \nodata & $63.3\pm1.5$ & $<70$ \\
\enddata
\tablenotetext{\dagger}{All uncertainties are statistical, and do not account for flux calibration uncertainties, which are typically on the order of 10\%.}
\end{deluxetable*}

%% file: table2.tex
\begin{deluxetable*}{ccc}
\tablenum{2}
\tablecaption{Summary of Analytic Model Parameters and Priors}
\label{table:analytic_priors}
\tablehead{\colhead{Parameter} & \colhead{Description} & \colhead{Prior}}
\startdata
$r_{c,r}$ & Central radius of the ring & $0.01'' < r_{c,r} < 0.5''$ \\ [2pt]
$r_{w,r}$ & Radial half-width of the ring & $-2 < \log_{10}{r_{w,r}} < \log_{10}{r_{c,r}}$ \\ [2pt]
$i$ & Inclination of the ring & $0^{\circ} < i < 90^{\circ}$ \\ [2pt]
$p.a.$ & Position angle of the ring & $0^{\circ} < r_{c,r} < 180^{\circ}$ \\ [2pt]
$F_{\nu,r}$ & Integrated flux of the ring & $-3 < \log_{10} F_{\nu,r} < 1$ \\ [2pt]
$r_{c,a}$ & Central radius of the asymmetry & $0.01'' < r_{c,r} < 0.5''$ \\ [2pt]
$r_{w,a}$ & Radial half-width of the asymmetry & $-2 < \log_{10}{r_{w,a}} < \log_{10}{r_{c,a}}$ \\ [2pt]
$\phi_{c,a}$ & Azimuthal center of the asymmetry & $0^{\circ} < \phi_{c,a} < 360^{\circ}$ \\ [2pt]
$\phi_{w,a}$ & Azimuthal half-width of the asymmetry & $0^{\circ} < \phi_{w,a} < 180^{\circ}$ \\ [2pt]
$F_{\nu,a}$ & Integrated flux of the asymmetry & $-3 < \log_{10} F_{\nu,a} < \log_{10} F_{\nu,r}$ \\ [2pt]
$r_{c,p}$ & Central radius of the point source & $0'' < r_{c,p} < \min(r_{c,r} - r_{w,r}, r_{c,a} - r_{w,a})$ \\ [2pt]
$r_{w,p}$ & Radial half-width of the point source & $-2.5 < \log_{10}{r_{w,p}} < \max(\log_{10}{r_{c,p}},-2)$ \\ [2pt]
$\phi_{c,p}$ & Azimuthal center of the point source & $0 < \log_{10}{\phi_{c,p}} < 180^{\circ}$ \\ [2pt]
$F_{\nu,p}$ & Integrated flux of the point source & $-3 < \log_{10} F_{\nu,p} < \log_{10} F_{\nu,a}$ \\ [2pt]
$r_{w,G}$ & 1$\sigma$ radius of the large scale Gaussian & $r_{c,r} < r_{w,G} < 3''$ \\ [2pt]
$F_{\nu,G}$ & Integrated flux of the large scale Gaussian & $-3 < \log_{10} F_{\nu,G} < 1$ \\ [2pt]
\enddata
\end{deluxetable*}

%% file: table3.tex
\begin{deluxetable*}{l|ccccccc}
\tablecaption{Best-fit Analytic Model Parameters}
\tablenum{3}
\tabletypesize{\normalsize}
\label{table:analytic_best_fits}
\tablehead{\colhead{Parameters} & \colhead{HH270MMS2} & \colhead{HOPS-56B} & \colhead{HOPS-65} & \colhead{HOPS-124} & \colhead{HOPS-140} & \colhead{HOPS-157} & \colhead{HOPS-163}}
\startdata
\multicolumn{8}{c}{Ring Component}\\[2pt]
\hline
$x_0$ (mas) & $-208.36^{+  1.25}_{-  1.50}$ & $582.95^{+ 12.71}_{-  9.35}$ & $-343.53^{+  3.25}_{-  4.13}$ & $1078.19^{+  0.22}_{-  0.19}$ & $215.18^{+ 17.15}_{- 21.63}$ & $-130.06^{+  7.61}_{-  9.14}$ & $ 46.88^{+  3.69}_{-  4.01}$ \\[2pt]
$y_0$ (mas) & $-383.21^{+  1.44}_{-  1.32}$ & $5553.80^{+  4.29}_{-  4.69}$ & $237.34^{+ 13.15}_{- 12.16}$ & $ 54.30^{+  0.21}_{-  0.20}$ & $191.94^{+  8.41}_{- 10.81}$ & $148.89^{+ 12.45}_{- 13.41}$ & $ 18.72^{+  2.89}_{-  2.89}$ \\[2pt]
$r_{c,r}$ (mas) & $105.0^{+  1.7}_{-  2.0}$ & $331.0^{+ 10.0}_{- 10.6}$ & $595.9^{+ 16.7}_{- 20.7}$ & $146.9^{+  0.7}_{-  0.6}$ & $408.5^{+ 30.3}_{- 20.4}$ & $225.4^{+ 13.7}_{- 17.0}$ & $296.6^{+  3.1}_{-  2.1}$ \\[2pt]
$r_{w,r}$ (mas) & $ 40.8^{+  2.6}_{-  3.0}$ & $114.6^{+ 12.2}_{-  7.8}$ & $291.2^{+ 19.2}_{- 19.9}$ & $ 90.3^{+  0.7}_{-  0.7}$ & $263.9^{+ 21.4}_{- 26.3}$ & $102.2^{+ 16.6}_{- 12.7}$ & $124.6^{+  5.3}_{-  4.8}$ \\[2pt]
$i$ ($^{\circ}$) & $34.15^{+ 2.23}_{- 2.10}$ & $58.96^{+ 0.91}_{- 1.17}$ & $75.64^{+ 0.51}_{- 0.52}$ & $45.91^{+ 0.12}_{- 0.12}$ & $56.77^{+ 1.73}_{- 2.26}$ & $39.94^{+ 4.27}_{- 4.71}$ & $67.53^{+ 0.55}_{- 1.06}$ \\[2pt]
$p.a.$ ($^{\circ}$) & $ 46.40^{+  4.26}_{-  3.20}$ & $ 57.94^{+  2.01}_{-  1.54}$ & $ -7.20^{+  0.43}_{-  0.53}$ & $  3.90^{+  0.14}_{-  0.14}$ & $ 71.54^{+  3.02}_{-  3.11}$ & $  4.89^{+  6.18}_{-  6.29}$ & $ 60.17^{+  0.32}_{-  1.00}$ \\[2pt]
$F_{\nu,r}$ (mJy) & $ 64.07^{+  1.43}_{-  1.35}$ & $ 62.06^{+  2.77}_{-  2.72}$ & $ 86.79^{+  2.99}_{-  2.94}$ & $941.21^{+  3.15}_{-  2.48}$ & $ 46.79^{+  3.21}_{-  4.04}$ & $ 25.45^{+  2.01}_{-  1.54}$ & $ 65.65^{+  1.43}_{-  1.24}$ \\[2pt]
\hline
\multicolumn{8}{c}{Asymmetry Component}\\[2pt]
\hline
$r_{c,a}$ (mas) & \nodata & $289.7^{+115.1}_{- 48.6}$ & $567.0^{+ 23.4}_{- 19.6}$ & $122.3^{+  0.9}_{-  0.8}$ & $314.4^{+ 19.1}_{- 23.8}$ & $241.4^{+ 55.3}_{- 23.9}$ & \nodata \\[2pt]
$r_{w,a}$ (mas) & \nodata & $147.4^{+ 83.0}_{- 67.1}$ & $ 88.1^{+ 17.3}_{- 21.1}$ & $ 28.3^{+  1.4}_{-  1.4}$ & $ 66.9^{+ 12.2}_{- 14.9}$ & $199.6^{+ 20.6}_{- 42.7}$ & \nodata \\[2pt]
$\phi_{c,a}$ ($^{\circ}$) & \nodata & $212.0^{+ 11.0}_{-  9.8}$ & $169.1^{+  6.4}_{-  7.0}$ & $ 93.5^{+  0.4}_{-  0.4}$ & $183.6^{+  7.0}_{-  5.7}$ & $171.8^{+  9.9}_{-  7.9}$ & \nodata \\[2pt]
$\phi_{w,a}$ ($^{\circ}$) & \nodata & $ 44.8^{+ 16.5}_{-  9.6}$ & $ 58.0^{+  8.3}_{-  8.2}$ & $ 56.8^{+  0.7}_{-  0.6}$ & $ 65.2^{+ 10.1}_{-  7.9}$ & $ 74.2^{+  9.1}_{-  6.8}$ & \nodata \\[2pt]
$F_{c,a}$ (mJy) & \nodata & $  6.02^{+  3.62}_{-  0.66}$ & $  9.98^{+  1.43}_{-  1.71}$ & $122.69^{+  1.57}_{-  1.86}$ & $ 13.10^{+  1.67}_{-  1.59}$ & $ 23.99^{+  1.39}_{-  1.91}$ & \nodata \\[2pt]
\hline
\multicolumn{8}{c}{Point Component}\\[2pt]
\hline
$\phi_{c,p}$ ($^{\circ}$) & \nodata & $-69.2^{+ 13.7}_{- 15.1}$ & $  4.4^{+ 19.2}_{- 16.1}$ & $-138.8^{+  3.2}_{-  3.7}$ & $ -5.9^{+ 11.9}_{- 20.3}$ & \nodata & \nodata \\[2pt]
$r_{c,p}$ (mas) & \nodata & $ 49.9^{+ 10.7}_{- 11.5}$ & $ 82.7^{+ 16.6}_{- 12.6}$ & $ 21.5^{+  1.0}_{-  1.2}$ & $120.3^{+ 24.5}_{- 24.8}$ & \nodata & \nodata \\[2pt]
$r_{w,p}$ (mas) & \nodata & $ 38.2^{+  8.8}_{-  7.4}$ & $ 45.5^{+ 11.5}_{- 12.3}$ & $  5.4^{+  1.7}_{-  0.4}$ & $ 16.9^{+ 19.1}_{- 11.1}$ & \nodata & \nodata \\[2pt]
$F_{c,p}$ (mJy) & \nodata & $ 5.47^{+ 0.68}_{- 0.52}$ & $ 5.26^{+ 0.76}_{- 0.89}$ & $22.83^{+ 0.88}_{- 1.03}$ & $ 1.17^{+ 0.42}_{- 0.15}$ & \nodata & \nodata \\[2pt]
\hline
\multicolumn{8}{c}{Gaussian Component}\\[2pt]
\hline
$\sigma_G$ (") & $ 0.334^{+ 0.074}_{- 0.061}$ & \nodata & \nodata & $ 0.725^{+ 0.017}_{- 0.015}$ & $ 1.215^{+ 0.745}_{- 0.342}$ & $ 1.076^{+ 0.281}_{- 0.209}$ & \nodata \\[2pt]
$F_G$ (mJy) & $ 15.3^{+  3.0}_{-  3.0}$ & \nodata & \nodata & $359.1^{+  7.3}_{-  8.6}$ & $ 25.7^{+ 14.8}_{- 11.2}$ & $ 40.0^{+ 11.3}_{-  8.8}$ & \nodata \\[2pt]
\hline
\multicolumn{8}{c}{log(Bayes Factor) = log(Model Bayesian Evidence) -- log(Full Model Bayesian Evidence)}\\[2pt]
\hline
Ring & $-52.4\pm{  0.3}$ & $-299.4\pm{  0.3}$ & $-342.6\pm{  0.3}$ & $-46042.8\pm{  0.4}$ & $-376.9\pm{  0.3}$ & $-237.8\pm{  0.3}$ & Full \\[2pt]
Ring+Gaussian & Full & \nodata & \nodata & \nodata & \nodata & \nodata & \nodata \\[2pt]
Asymmetric Ring & \nodata & $-127.3\pm{  0.3}$ & $-136.1\pm{  0.3}$ & $-9546.1\pm{  0.4}$ & $-15.2\pm{  0.3}$ & $-59.8\pm{  0.3}$ & \nodata \\[2pt]
Asymmetric Ring+\\Point & \nodata & Full & Full & $-9234.9\pm{  0.4}$ & $-11.8\pm{  0.3}$ & \nodata & \nodata \\[2pt]
Asymmetric Ring\\+Gaussian & \nodata & \nodata & \nodata & $-915.6\pm{  0.4}$ & $ -4.2\pm{  0.3}$ & Full & \nodata \\[2pt]
Asymmetric Ring\\+Point+Gaussian & \nodata & \nodata & \nodata & Full & Full & \nodata & \nodata
\enddata
\end{deluxetable*}

%% file: table4.tex
\begin{deluxetable*}{l|ccccccc}
\tablecaption{Best-fit Radiative Transfer Model Parameters}
\tablenum{4}
\tabletypesize{\normalsize}
\label{table:rt_best_fits}
\tablehead{\colhead{Parameters} & \colhead{HH270MMS2} & \colhead{HOPS-56B} & \colhead{HOPS-65} & \colhead{HOPS-124} & \colhead{HOPS-140} & \colhead{HOPS-157} & \colhead{HOPS-163}}
\startdata
\multicolumn{8}{c}{Star}\\[2pt]
\hline
$L_*$ (L$_{\odot}$) & $3.01^{+0.50}_{-0.44}$ & $187.89^{+120.31}_{-123.70}$ & $0.97^{+0.39}_{-0.10}$ & $243.09^{+15.87}_{-30.09}$ & $0.41^{+0.06}_{-0.06}$ & $1.96^{+0.19}_{-0.15}$ & $1.17^{+0.11}_{-0.10}$ \\[2pt]
\hline
\multicolumn{8}{c}{Disk}\\[2pt]
\hline
$M_{disk}$ (M$_{\odot}$) & $0.026^{+0.006}_{-0.004}$ & $0.004^{+0.002}_{-0.001}$ & $0.399^{+0.415}_{-0.203}$ & $0.093^{+0.013}_{-0.010}$ & $0.139^{+0.014}_{-0.124}$ & $0.021^{+0.025}_{-0.014}$ & $0.040^{+0.071}_{-0.015}$ \\[2pt]
$R_{disk}$ (au) & $ 40.8^{+  1.1}_{-  1.8}$ & $125.1^{+ 11.7}_{-  8.4}$ & $183.8^{+ 20.1}_{- 13.9}$ & $ 44.5^{+  3.7}_{-  2.9}$ & $215.8^{+ 12.8}_{- 14.6}$ & $116.8^{+ 10.0}_{-  5.9}$ & $133.1^{+ 17.5}_{- 18.6}$ \\[2pt]
$\gamma$ & $-0.4^{+ 0.1}_{- 0.0}$ & $-0.5^{+ 0.2}_{- 0.0}$ & $ 0.1^{+ 0.1}_{- 0.1}$ & $ 0.4^{+ 0.1}_{- 0.1}$ & $-0.4^{+ 0.3}_{- 0.1}$ & $-0.4^{+ 0.2}_{- 0.1}$ & $-0.3^{+ 0.7}_{- 0.2}$ \\[2pt]
$h_{0, 1\,\mathrm{au}}$ (au) & $0.03^{+0.01}_{-0.01}$ & $0.16^{+0.12}_{-0.11}$ & $0.39^{+0.03}_{-0.05}$ & $0.15^{+0.03}_{-0.01}$ & $0.10^{+0.08}_{-0.05}$ & $0.33^{+0.31}_{-0.09}$ & $0.05^{+0.04}_{-0.03}$ \\[2pt]
$\beta$ & $0.81^{+0.18}_{-0.25}$ & $1.02^{+0.25}_{-0.12}$ & $0.71^{+0.03}_{-0.02}$ & $0.92^{+0.02}_{-0.02}$ & $1.08^{+0.05}_{-0.47}$ & $0.53^{+0.08}_{-0.03}$ & $0.56^{+0.26}_{-0.06}$ \\[2pt]
$R_{cav}$ (au) & $ 33.2^{+  0.5}_{-  1.2}$ & \nodata & \nodata & \nodata & \nodata & $ 57.3^{+  6.6}_{-  7.1}$ & $ 84.4^{+  5.5}_{-  4.4}$ \\[2pt]
$\delta_{cav}$ & $0.033^{+0.013}_{-0.013}$ & \nodata & \nodata & \nodata & \nodata & $0.185^{+0.103}_{-0.178}$ & $0.005^{+0.010}_{-0.004}$ \\[2pt]
$R_{gap}$ (au) & \nodata & $ 64.6^{+  2.3}_{-  3.3}$ & $110.8^{+  4.0}_{-  9.1}$ & $ 23.7^{+  1.4}_{-  1.4}$ & $ 51.0^{+  4.7}_{-  4.9}$ & \nodata & \nodata \\[2pt]
$w_{gap}$ (au) & \nodata & $ 97.0^{+ 10.9}_{-  8.7}$ & $216.3^{+  9.5}_{- 11.3}$ & $ 33.8^{+  2.6}_{-  1.4}$ & $ 90.8^{+ 11.0}_{- 12.8}$ & \nodata & \nodata \\[2pt]
$\delta_{gap}$ & \nodata & $0.025^{+0.019}_{-0.025}$ & $0.008^{+0.007}_{-0.003}$ & $0.003^{+0.002}_{-0.001}$ & $0.074^{+0.064}_{-0.074}$ & \nodata & \nodata \\[2pt]
\hline
\multicolumn{8}{c}{Envelope}\\[2pt]
\hline
$M_{env}$ (M$_{\odot}$) & $0.0060^{+0.0007}_{-0.0007}$ & $0.0029^{+0.0231}_{-0.0029}$ & $0.0013^{+0.0008}_{-0.0003}$ & $0.1793^{+0.0306}_{-0.0270}$ & $0.0155^{+0.0292}_{-0.0097}$ & $1.8761^{+0.8877}_{-0.5451}$ & $0.0014^{+0.0010}_{-0.0005}$ \\[2pt]
$R_{env}$ (au) & $ 519.2^{+  17.5}_{-  21.6}$ & $1245.4^{+5525.5}_{-1038.8}$ & $ 468.1^{+ 233.9}_{-  57.6}$ & $1308.0^{+ 239.6}_{- 198.3}$ & $1387.7^{+1730.4}_{- 625.5}$ & $14356.7^{+4672.0}_{-2942.1}$ & $ 246.6^{+ 135.2}_{-  21.7}$ \\[2pt]
$\xi$ & $1.076^{+0.012}_{-0.013}$ & $1.277^{+0.205}_{-0.620}$ & $0.981^{+0.501}_{-0.459}$ & $1.249^{+0.020}_{-0.110}$ & $1.224^{+0.264}_{-0.701}$ & $1.404^{+0.085}_{-0.275}$ & $0.533^{+0.089}_{-0.032}$ \\[2pt]
$f_{cav}$ & $0.12^{+0.02}_{-0.02}$ & $0.78^{+0.20}_{-0.70}$ & $0.82^{+0.16}_{-0.79}$ & $0.13^{+0.46}_{-0.09}$ & $0.83^{+0.16}_{-0.68}$ & $0.47^{+0.07}_{-0.06}$ & $0.08^{+0.08}_{-0.04}$ \\[2pt]
\hline
\multicolumn{8}{c}{Dust}\\[2pt]
\hline
$a_{max}$ ($\mu$m) & $   61^{+   16}_{-   21}$ & $ 1955^{+ 3121}_{- 1158}$ & $34452^{+57756}_{-20246}$ & $ 9024^{+ 6118}_{- 2926}$ & $ 7942^{+78124}_{- 7938}$ & $19476^{+71967}_{-15076}$ & $10198^{+78611}_{- 7449}$ \\[2pt]
$p$ & $2.92^{+0.32}_{-0.22}$ & $3.42^{+0.27}_{-0.37}$ & $2.65^{+0.26}_{-0.13}$ & $3.86^{+0.13}_{-0.10}$ & $4.32^{+0.17}_{-1.09}$ & $2.83^{+0.39}_{-0.31}$ & $3.15^{+1.17}_{-0.59}$ \\[2pt]
\hline
\multicolumn{8}{c}{Viewing}\\[2pt]
\hline
$i$ ($^{\circ}$) & $33.8^{+ 1.4}_{- 1.1}$ & $61.8^{+ 1.4}_{- 1.1}$ & $77.7^{+ 0.4}_{- 1.3}$ & $44.8^{+ 1.5}_{- 1.2}$ & $56.9^{+ 2.2}_{- 2.3}$ & $ 2.8^{+18.8}_{- 2.7}$ & $68.1^{+ 0.8}_{- 0.9}$ \\[2pt]
p.a. ($^{\circ}$) & $138.2^{+  2.0}_{-  1.7}$ & $150.7^{+  1.2}_{-  1.3}$ & $ 82.7^{+  1.6}_{-  0.4}$ & $ 94.4^{+  2.8}_{-  2.9}$ & $163.6^{+  2.8}_{-  2.6}$ & $126.0^{+ 50.7}_{- 70.2}$ & $150.1^{+  0.9}_{-  0.9}$
\enddata
\end{deluxetable*}

%% file: ms.bbl
\begin{thebibliography}{}
\expandafter\ifx\csname natexlab\endcsname\relax\def\natexlab#1{#1}\fi

\bibitem[{{Alexander} {et~al.}(2006){Alexander}, {Clarke}, \&
  {Pringle}}]{Alexander2006}
{Alexander}, R.~D., {Clarke}, C.~J., \& {Pringle}, J.~E. 2006, \mnras, 369, 216

\bibitem[{{ALMA Partnership} {et~al.}(2015){ALMA Partnership}, {Brogan},
  {P{\'e}rez}, {Hunter}, {Dent}, {Hales}, {Hills}, {Corder}, {Fomalont},
  {Vlahakis}, {Asaki}, {Barkats}, {Hirota}, {Hodge}, {Impellizzeri}, {Kneissl},
  {Liuzzo}, {Lucas}, {Marcelino}, {Matsushita}, {Nakanishi}, {Phillips},
  {Richards}, {Toledo}, {Aladro}, {Broguiere}, {Cortes}, {Cortes}, {Espada},
  {Galarza}, {Garcia-Appadoo}, {Guzman-Ramirez}, {Humphreys}, {Jung}, {Kameno},
  {Laing}, {Leon}, {Marconi}, {Mignano}, {Nikolic}, {Nyman}, {Radiszcz},
  {Remijan}, {Rod{\'o}n}, {Sawada}, {Takahashi}, {Tilanus}, {Vila Vilaro},
  {Watson}, {Wiklind}, {Akiyama}, {Chapillon}, {de Gregorio-Monsalvo}, {Di
  Francesco}, {Gueth}, {Kawamura}, {Lee}, {Nguyen Luong}, {Mangum}, {Pietu},
  {Sanhueza}, {Saigo}, {Takakuwa}, {Ubach}, {van Kempen}, {Wootten},
  {Castro-Carrizo}, {Francke}, {Gallardo}, {Garcia}, {Gonzalez}, {Hill},
  {Kaminski}, {Kurono}, {Liu}, {Lopez}, {Morales}, {Plarre}, {Schieven},
  {Testi}, {Videla}, {Villard}, {Andreani}, {Hibbard}, \&
  {Tatematsu}}]{Brogan2015}
{ALMA Partnership}, {Brogan}, C.~L., {P{\'e}rez}, L.~M., {et~al.} 2015, \apjl,
  808, L3

\bibitem[{{Andersen} {et~al.}(2019){Andersen}, {Stephens}, {Dunham}, {Pokhrel},
  {J{\o}rgensen}, {Frimann}, {Segura-Cox}, {Myers}, {Bourke}, {Tobin}, \&
  {Tychoniec}}]{Andersen2019}
{Andersen}, B.~C., {Stephens}, I.~W., {Dunham}, M.~M., {et~al.} 2019, \apj,
  873, 54

\bibitem[{{Andrews} {et~al.}(2016){Andrews}, {Wilner}, {Zhu}, {Birnstiel},
  {Carpenter}, {P{\'e}rez}, {Bai}, {{\"O}berg}, {Hughes}, {Isella}, \&
  {Ricci}}]{Andrews2016}
{Andrews}, S.~M., {Wilner}, D.~J., {Zhu}, Z., {et~al.} 2016, \apjl, 820, L40

\bibitem[{{Andrews} {et~al.}(2018){Andrews}, {Huang}, {P{\'e}rez}, {Isella},
  {Dullemond}, {Kurtovic}, {Guzm{\'a}n}, {Carpenter}, {Wilner}, {Zhang}, {Zhu},
  {Birnstiel}, {Bai}, {Benisty}, {Hughes}, {{\"O}berg}, \&
  {Ricci}}]{Andrews2018}
{Andrews}, S.~M., {Huang}, J., {P{\'e}rez}, L.~M., {et~al.} 2018, \apjl, 869,
  L41

\bibitem[{{Armitage}(2011)}]{Armitage2011}
{Armitage}, P.~J. 2011, \araa, 49, 195

\bibitem[{{Bate}(2018)}]{Bate2018}
{Bate}, M.~R. 2018, \mnras, 475, 5618

\bibitem[{{Birnstiel} {et~al.}(2012){Birnstiel}, {Andrews}, \&
  {Ercolano}}]{Birnstiel2012}
{Birnstiel}, T., {Andrews}, S.~M., \& {Ercolano}, B. 2012, \aap, 544, A79

\bibitem[{{Boss}(1997)}]{Boss1997}
{Boss}, A.~P. 1997, Science, 276, 1836

\bibitem[{{Boss}(2011)}]{Boss2011}
---. 2011, \apj, 731, 74

\bibitem[{{Calcino} {et~al.}(2019){Calcino}, {Price}, {Pinte}, {van der Marel},
  {Ragusa}, {Dipierro}, {Cuello}, \& {Christiaens}}]{Calcino2019}
{Calcino}, J., {Price}, D.~J., {Pinte}, C., {et~al.} 2019, arXiv e-prints,
  arXiv:1910.00161

\bibitem[{{Chen} {et~al.}(1995){Chen}, {Myers}, {Ladd}, \& {Wood}}]{Chen1995}
{Chen}, H., {Myers}, P.~C., {Ladd}, E.~F., \& {Wood}, D.~O.~S. 1995, \apj, 445,
  377

\bibitem[{{Chiang} \& {Goldreich}(1999)}]{Chiang1999}
{Chiang}, E.~I., \& {Goldreich}, P. 1999, \apj, 519, 279

\bibitem[{{Clarke} {et~al.}(2001){Clarke}, {Gendrin}, \&
  {Sotomayor}}]{Clarke2001}
{Clarke}, C.~J., {Gendrin}, A., \& {Sotomayor}, M. 2001, \mnras, 328, 485

\bibitem[{{Crapsi} {et~al.}(2008){Crapsi}, {van Dishoeck}, {Hogerheijde},
  {Pontoppidan}, \& {Dullemond}}]{Crapsi2008}
{Crapsi}, A., {van Dishoeck}, E.~F., {Hogerheijde}, M.~R., {Pontoppidan},
  K.~M., \& {Dullemond}, C.~P. 2008, \aap, 486, 245

\bibitem[{{de Valon} {et~al.}(2020){de Valon}, {Dougados}, {Cabrit}, {Louvet},
  {Zapata}, \& {Mardones}}]{deValon2020}
{de Valon}, A., {Dougados}, C., {Cabrit}, S., {et~al.} 2020, arXiv e-prints,
  arXiv:2001.09776

\bibitem[{{Dodson-Robinson} \& {Salyk}(2011)}]{DodsonRobinson2011}
{Dodson-Robinson}, S.~E., \& {Salyk}, C. 2011, \apj, 738, 131

\bibitem[{{Dong} {et~al.}(2015){Dong}, {Zhu}, \& {Whitney}}]{Dong2015}
{Dong}, R., {Zhu}, Z., \& {Whitney}, B. 2015, \apj, 809, 93

\bibitem[{{Dr{\c{a}}{\.z}kowska} \& {Dullemond}(2018)}]{Drazkowska2018}
{Dr{\c{a}}{\.z}kowska}, J., \& {Dullemond}, C.~P. 2018, \aap, 614, A62

\bibitem[{{Dullemond}(2012)}]{Dullemond2012}
{Dullemond}, C.~P. 2012, {RADMC-3D: A multi-purpose radiative transfer tool},
  Astrophysics Source Code Library, , , ascl:1202.015

\bibitem[{{Dullemond} \& {Dominik}(2005)}]{Dullemond2005}
{Dullemond}, C.~P., \& {Dominik}, C. 2005, \aap, 434, 971

\bibitem[{{Dunham} {et~al.}(2015){Dunham}, {Allen}, {Evans},
  {Broekhoven-Fiene}, {Cieza}, {Di Francesco}, {Gutermuth}, {Harvey},
  {Hatchell}, {Heiderman}, {Huard}, {Johnstone}, {Kirk}, {Matthews}, {Miller},
  {Peterson}, \& {Young}}]{Dunham2015}
{Dunham}, M.~M., {Allen}, L.~E., {Evans}, II, N.~J., {et~al.} 2015, \apjs, 220,
  11

\bibitem[{{Durisen} {et~al.}(2007){Durisen}, {Boss}, {Mayer}, {Nelson},
  {Quinn}, \& {Rice}}]{Durisen2007}
{Durisen}, R.~H., {Boss}, A.~P., {Mayer}, L., {et~al.} 2007, in Protostars and
  Planets V, ed. B.~{Reipurth}, D.~{Jewitt}, \& K.~{Keil}, 607

\bibitem[{{Evans} {et~al.}(2009){Evans}, {Dunham}, {J{\o}rgensen}, {Enoch},
  {Mer{\'{\i}}n}, {van Dishoeck}, {Alcal{\'a}}, {Myers}, {Stapelfeldt},
  {Huard}, {Allen}, {Harvey}, {van Kempen}, {Blake}, {Koerner}, {Mundy},
  {Padgett}, \& {Sargent}}]{Evans2009}
{Evans}, II, N.~J., {Dunham}, M.~M., {J{\o}rgensen}, J.~K., {et~al.} 2009,
  \apjs, 181, 321

\bibitem[{{Fischer} {et~al.}(2019){Fischer}, {Safron}, \&
  {Megeath}}]{Fischer2019}
{Fischer}, W.~J., {Safron}, E., \& {Megeath}, S.~T. 2019, \apj, 872, 183

\bibitem[{{Fischer} {et~al.}(2013){Fischer}, {Megeath}, {Stutz}, {Tobin},
  {Ali}, {Stanke}, {Osorio}, {Furlan}, {HOPS Team}, \& {Orion Protostar
  Survey}}]{Fischer2013}
{Fischer}, W.~J., {Megeath}, S.~T., {Stutz}, A.~M., {et~al.} 2013,
  Astronomische Nachrichten, 334, 53

\bibitem[{{Flock} {et~al.}(2015){Flock}, {Ruge}, {Dzyurkevich}, {Henning},
  {Klahr}, \& {Wolf}}]{Flock2015}
{Flock}, M., {Ruge}, J.~P., {Dzyurkevich}, N., {et~al.} 2015, \aap, 574, A68

\bibitem[{Foreman-Mackey(2016)}]{ForemanMackey2016}
Foreman-Mackey, D. 2016, The Journal of Open Source Software, 24,
  doi:10.21105/joss.00024

\bibitem[{{Foreman-Mackey} {et~al.}(2013){Foreman-Mackey}, {Hogg}, {Lang}, \&
  {Goodman}}]{ForemanMackey2013}
{Foreman-Mackey}, D., {Hogg}, D.~W., {Lang}, D., \& {Goodman}, J. 2013, \pasp,
  125, 306

\bibitem[{{Forgan} \& {Rice}(2013)}]{Forgan2013}
{Forgan}, D., \& {Rice}, K. 2013, \mnras, 432, 3168

\bibitem[{{Furlan} {et~al.}(2008){Furlan}, {McClure}, {Calvet}, {Hartmann},
  {D'Alessio}, {Forrest}, {Watson}, {Uchida}, {Sargent}, {Green}, \&
  {Herter}}]{Furlan2008}
{Furlan}, E., {McClure}, M., {Calvet}, N., {et~al.} 2008, \apjs, 176, 184

\bibitem[{{Furlan} {et~al.}(2016){Furlan}, {Fischer}, {Ali}, {Stutz}, {Stanke},
  {Tobin}, {Megeath}, {Osorio}, {Hartmann}, {Calvet}, {Poteet}, {Booker},
  {Manoj}, {Watson}, \& {Allen}}]{Furlan2016}
{Furlan}, E., {Fischer}, W.~J., {Ali}, B., {et~al.} 2016, \apjs, 224, 5

\bibitem[{{Gaia Collaboration} {et~al.}(2018){Gaia Collaboration}, {Brown},
  {Vallenari}, {Prusti}, {de Bruijne}, {Babusiaux}, {Bailer-Jones}, {Biermann},
  {Evans}, {Eyer}, \& et~al.}]{GaiaCollaboration2018}
{Gaia Collaboration}, {Brown}, A.~G.~A., {Vallenari}, A., {et~al.} 2018, \aap,
  616, A1

\bibitem[{{Gorti} \& {Hollenbach}(2009)}]{Gorti2009a}
{Gorti}, U., \& {Hollenbach}, D. 2009, \apj, 690, 1539

\bibitem[{{Hildebrand}(1983)}]{Hildebrand1983}
{Hildebrand}, R.~H. 1983, \qjras, 24, 267

\bibitem[{{Huang} {et~al.}(2018{\natexlab{a}}){Huang}, {Andrews}, {Dullemond},
  {Isella}, {P{\'e}rez}, {Guzm{\'a}n}, {{\"O}berg}, {Zhu}, {Zhang}, {Bai},
  {Benisty}, {Birnstiel}, {Carpenter}, {Hughes}, {Ricci}, {Weaver}, \&
  {Wilner}}]{Huang2018}
{Huang}, J., {Andrews}, S.~M., {Dullemond}, C.~P., {et~al.} 2018{\natexlab{a}},
  \apjl, 869, L42

\bibitem[{{Huang} {et~al.}(2018{\natexlab{b}}){Huang}, {Andrews}, {P{\'e}rez},
  {Zhu}, {Dullemond}, {Isella}, {Benisty}, {Bai}, {Birnstiel}, {Carpenter},
  {Guzm{\'a}n}, {Hughes}, {{\"O}berg}, {Ricci}, {Wilner}, \&
  {Zhang}}]{Huang2018b}
{Huang}, J., {Andrews}, S.~M., {P{\'e}rez}, L.~M., {et~al.} 2018{\natexlab{b}},
  \apjl, 869, L43

\bibitem[{Hunter(2007)}]{Hunter2007}
Hunter, J.~D. 2007, Computing In Science \& Engineering, 9, 90

\bibitem[{{Isella} {et~al.}(2016){Isella}, Guidi, Testi, Liu, Li, Li, Weaver,
  Boehler, Carperter, De~Gregorio-Monsalvo, Manara, Natta, P\'erez, Ricci,
  Sargent, Tazzari, \& Turner}]{Isella2016}
{Isella}, A., Guidi, G., Testi, L., {et~al.} 2016, Phys. Rev. Lett., 117,
  251101

\bibitem[{{Kounkel} {et~al.}(2016){Kounkel}, {Megeath}, {Poteet}, {Fischer}, \&
  {Hartmann}}]{Kounkel2016}
{Kounkel}, M., {Megeath}, S.~T., {Poteet}, C.~A., {Fischer}, W.~J., \&
  {Hartmann}, L. 2016, \apj, 821, 52

\bibitem[{{Kounkel} {et~al.}(2017){Kounkel}, {Hartmann}, {Loinard},
  {Ortiz-Le{\'o}n}, {Mioduszewski}, {Rodr{\'\i}guez}, {Dzib}, {Torres}, {Pech},
  {Galli}, {Rivera}, {Boden}, {Evans}, {Brice{\~n}o}, \& {Tobin}}]{Kounkel2017}
{Kounkel}, M., {Hartmann}, L., {Loinard}, L., {et~al.} 2017, \apj, 834, 142

\bibitem[{{Kounkel} {et~al.}(2018){Kounkel}, {Covey}, {Su{\'a}rez},
  {Rom{\'a}n-Z{\'u}{\~n}iga}, {Hernandez}, {Stassun}, {Jaehnig}, {Feigelson},
  {Pe{\~n}a Ram{\'\i}rez}, {Roman-Lopes}, {Da Rio}, {Stringfellow}, {Kim},
  {Borissova}, {Fern{\'a}ndez-Trincado}, {Burgasser},
  {Garc{\'\i}a-Hern{\'a}ndez}, {Zamora}, {Pan}, \& {Nitschelm}}]{Kounkel2018}
{Kounkel}, M., {Covey}, K., {Su{\'a}rez}, G., {et~al.} 2018, \aj, 156, 84

\bibitem[{{Kratter} {et~al.}(2010){Kratter}, {Murray-Clay}, \&
  {Youdin}}]{Kratter2010}
{Kratter}, K.~M., {Murray-Clay}, R.~A., \& {Youdin}, A.~N. 2010, \apj, 710,
  1375

\bibitem[{{Lada}(1987)}]{Lada1987}
{Lada}, C.~J. 1987, in IAU Symposium, Vol. 115, Star Forming Regions, ed.
  M.~{Peimbert} \& J.~{Jugaku}, 1--17

\bibitem[{{Lee} {et~al.}(2019){Lee}, {Li}, \& {Turner}}]{Lee2019b}
{Lee}, C.-F., {Li}, Z.-Y., \& {Turner}, N.~J. 2019, Nature Astronomy, 466

\bibitem[{{Long} {et~al.}(2018){Long}, {Pinilla}, {Herczeg}, {Harsono},
  {Dipierro}, {Pascucci}, {Hendler}, {Tazzari}, {Ragusa}, {Salyk}, {Edwards},
  {Lodato}, {van de Plas}, {Johnstone}, {Liu}, {Boehler}, {Cabrit}, {Manara},
  {Menard}, {Mulders}, {Nisini}, {Fischer}, {Rigliaco}, {Banzatti}, {Avenhaus},
  \& {Gully-Santiago}}]{Long2018}
{Long}, F., {Pinilla}, P., {Herczeg}, G.~J., {et~al.} 2018, \apj, 869, 17

\bibitem[{{Lynden-Bell} \& {Pringle}(1974)}]{LyndenBell1974}
{Lynden-Bell}, D., \& {Pringle}, J.~E. 1974, \mnras, 168, 603

\bibitem[{{Manara} {et~al.}(2018){Manara}, {Morbidelli}, \&
  {Guillot}}]{Manara2018}
{Manara}, C.~F., {Morbidelli}, A., \& {Guillot}, T. 2018, \aap, 618, L3

\bibitem[{{McMullin} {et~al.}(2007){McMullin}, {Waters}, {Schiebel}, {Young},
  \& {Golap}}]{McMullin2007}
{McMullin}, J.~P., {Waters}, B., {Schiebel}, D., {Young}, W., \& {Golap}, K.
  2007, in Astronomical Society of the Pacific Conference Series, Vol. 376,
  Astronomical Data Analysis Software and Systems XVI, ed. R.~A. {Shaw},
  F.~{Hill}, \& D.~J. {Bell}, 127

\bibitem[{{Myers} {et~al.}(1987){Myers}, {Fuller}, {Mathieu}, {Beichman},
  {Benson}, {Schild}, \& {Emerson}}]{Myers1987}
{Myers}, P.~C., {Fuller}, G.~A., {Mathieu}, R.~D., {et~al.} 1987, \apj, 319,
  340

\bibitem[{{Najita} \& {Kenyon}(2014)}]{Najita2014}
{Najita}, J.~R., \& {Kenyon}, S.~J. 2014, \mnras, 445, 3315

\bibitem[{{Okuzumi} {et~al.}(2016){Okuzumi}, {Momose}, {Sirono}, {Kobayashi},
  \& {Tanaka}}]{Okuzumi2016}
{Okuzumi}, S., {Momose}, M., {Sirono}, S.-i., {Kobayashi}, H., \& {Tanaka}, H.
  2016, \apj, 821, 82

\bibitem[{{Ossenkopf} \& {Henning}(1994)}]{Ossenkopf1994}
{Ossenkopf}, V., \& {Henning}, T. 1994, \aap, 291, 943

\bibitem[{{Owen} {et~al.}(2010){Owen}, {Ercolano}, {Clarke}, \&
  {Alexander}}]{Owen2010}
{Owen}, J.~E., {Ercolano}, B., {Clarke}, C.~J., \& {Alexander}, R.~D. 2010,
  \mnras, 401, 1415

\bibitem[{{P{\'e}rez} {et~al.}(2016){P{\'e}rez}, Carpenter, Andrews, Ricci,
  Isella, Linz, Sargent, Wilner, Henning, Deller, Chandler, Dullemond, Lazio,
  Menten, Corder, Storm, Testi, Tazzari, Kwon, Calvet, Greaves, Harris, \&
  Mundy}]{Perez2016}
{P{\'e}rez}, L.~M., Carpenter, J.~M., Andrews, S.~M., {et~al.} 2016, Science,
  353, 1519

\bibitem[{{Pinilla} {et~al.}(2012){Pinilla}, {Birnstiel}, {Ricci}, {Dullemond},
  {Uribe}, {Testi}, \& {Natta}}]{Pinilla2012}
{Pinilla}, P., {Birnstiel}, T., {Ricci}, L., {et~al.} 2012, \aap, 538, A114

\bibitem[{{Poblete} {et~al.}(2019){Poblete}, {Cuello}, \&
  {Cuadra}}]{Poblete2019}
{Poblete}, P.~P., {Cuello}, N., \& {Cuadra}, J. 2019, \mnras, 489, 2204

\bibitem[{{Price} {et~al.}(2018){Price}, {Cuello}, {Pinte}, {Mentiplay},
  {Casassus}, {Christiaens}, {Kennedy}, {Cuadra}, {Sebastian Perez}, {Marino},
  {Armitage}, {Zurlo}, {Juhasz}, {Ragusa}, {Laibe}, \& {Lodato}}]{Price2018}
{Price}, D.~J., {Cuello}, N., {Pinte}, C., {et~al.} 2018, \mnras, 477, 1270

\bibitem[{{Robitaille} {et~al.}(2006){Robitaille}, {Whitney}, {Indebetouw},
  {Wood}, \& {Denzmore}}]{Robitaille2006}
{Robitaille}, T.~P., {Whitney}, B.~A., {Indebetouw}, R., {Wood}, K., \&
  {Denzmore}, P. 2006, \apjs, 167, 256

\bibitem[{{Safron} {et~al.}(2015){Safron}, {Fischer}, {Megeath}, {Furlan},
  {Stutz}, {Stanke}, {Billot}, {Rebull}, {Tobin}, {Ali}, {Allen}, {Booker},
  {Watson}, \& {Wilson}}]{Safron2015}
{Safron}, E.~J., {Fischer}, W.~J., {Megeath}, S.~T., {et~al.} 2015, \apjl, 800,
  L5

\bibitem[{Sheehan(2018)}]{Sheehan2018b}
Sheehan, P. 2018, {psheehan/pdspy: pdspy: A MCMC Tool for Continuum and
  Spectral Line Radiative Transfer Modeling}, , , doi:10.5281/zenodo.2455079

\bibitem[{{Sheehan} \& {Eisner}(2017{\natexlab{a}})}]{Sheehan2017b}
{Sheehan}, P.~D., \& {Eisner}, J.~A. 2017{\natexlab{a}}, \apj, 851, 45

\bibitem[{{Sheehan} \& {Eisner}(2017{\natexlab{b}})}]{Sheehan2017a}
---. 2017{\natexlab{b}}, \apjl, 840, L12

\bibitem[{{Sheehan} \& {Eisner}(2018)}]{Sheehan2018}
---. 2018, \apj, 857, 18

\bibitem[{{Sheehan} {et~al.}(2019){Sheehan}, {Wu}, {Eisner}, \&
  {Tobin}}]{Sheehan2019}
{Sheehan}, P.~D., {Wu}, Y.-L., {Eisner}, J.~A., \& {Tobin}, J.~J. 2019, \apj,
  874, 136

\bibitem[{{Speagle}(2019)}]{Speagle2019a}
{Speagle}, J.~S. 2019, arXiv e-prints, arXiv:1904.02180

\bibitem[{{Takahashi} \& {Muto}(2018)}]{Takahashi2018}
{Takahashi}, S.~Z., \& {Muto}, T. 2018, \apj, 865, 102

\bibitem[{{Tazzari} {et~al.}(2018){Tazzari}, {Beaujean}, \&
  {Testi}}]{Tazzari2018}
{Tazzari}, M., {Beaujean}, F., \& {Testi}, L. 2018, \mnras, 476, 4527

\bibitem[{{Tazzari} {et~al.}(2017){Tazzari}, {Testi}, {Natta}, {Ansdell},
  {Carpenter}, {Guidi}, {Hogerheijde}, {Manara}, {Miotello}, {van der Marel},
  {van Dishoeck}, \& {Williams}}]{Tazzari2017a}
{Tazzari}, M., {Testi}, L., {Natta}, A., {et~al.} 2017, \aap, 606, A88

\bibitem[{{Tobin} {et~al.}(2016){Tobin}, {Kratter}, {Persson}, {Looney},
  {Dunham}, {Segura-Cox}, {Li}, {Chandler}, {Sadavoy}, {Harris}, {Melis}, \&
  {P{\'e}rez}}]{Tobin2016b}
{Tobin}, J.~J., {Kratter}, K.~M., {Persson}, M.~V., {et~al.} 2016, \nat, 538,
  483

\bibitem[{{Tobin} {et~al.}(2020){Tobin}, {Sheehan}, {Megeath},
  {D{\'\i}az-Rodr{\'\i}guez}, {Offner}, {Murillo}, {van 't Hoff}, {van
  Dishoeck}, {Osorio}, {Anglada}, {Furlan}, {Stutz}, {Reynolds}, {Karnath},
  {Fischer}, {Persson}, {Looney}, {Li}, {Stephens}, {Chand ler}, {Cox},
  {Dunham}, {Tychoniec}, {Kama}, {Kratter}, {Kounkel}, {Mazur}, {Maud},
  {Patel}, {Perez}, {Sadavoy}, {Segura-Cox}, {Sharma}, {Stephenson}, {Watson},
  \& {Wyrowski}}]{Tobin2020}
{Tobin}, J.~J., {Sheehan}, P.~D., {Megeath}, S.~T., {et~al.} 2020, \apj, 890,
  130

\bibitem[{{Ulrich}(1976)}]{Ulrich1976}
{Ulrich}, R.~K. 1976, \apj, 210, 377

\bibitem[{{van der Marel} {et~al.}(2019){van der Marel}, {Dong}, {di
  Francesco}, {Williams}, \& {Tobin}}]{vanderMarel2019}
{van der Marel}, N., {Dong}, R., {di Francesco}, J., {Williams}, J.~P., \&
  {Tobin}, J. 2019, \apj, 872, 112

\bibitem[{{van der Marel} {et~al.}(2013){van der Marel}, {Kristensen},
  {Visser}, {Mottram}, {Y{\i}ld{\i}z}, \& {van Dishoeck}}]{vanderMarel2013}
{van der Marel}, N., {Kristensen}, L.~E., {Visser}, R., {et~al.} 2013, \aap,
  556, A76

\bibitem[{{van der Marel} {et~al.}(2015){van der Marel}, {Pinilla}, {Tobin},
  {van Kempen}, {Andrews}, {Ricci}, \& {Birnstiel}}]{vanderMarel2015}
{van der Marel}, N., {Pinilla}, P., {Tobin}, J., {et~al.} 2015, \apjl, 810, L7

\bibitem[{{Vorobyov} \& {Basu}(2010)}]{Vorobyov2010}
{Vorobyov}, E.~I., \& {Basu}, S. 2010, \apj, 719, 1896

\bibitem[{{Wachmann}(1939)}]{Wachmann1939}
{Wachmann}, A.~A. 1939, IAU Circ., 738

\bibitem[{{Weidenschilling}(1977)}]{Weidenschilling1977b}
{Weidenschilling}, S.~J. 1977, \mnras, 180, 57

\bibitem[{{Zhang} {et~al.}(2015){Zhang}, {Blake}, \& {Bergin}}]{Zhang2015}
{Zhang}, K., {Blake}, G.~A., \& {Bergin}, E.~A. 2015, \apjl, 806, L7

\bibitem[{{Zhu} {et~al.}(2012){Zhu}, {Hartmann}, {Nelson}, \&
  {Gammie}}]{Zhu2012}
{Zhu}, Z., {Hartmann}, L., {Nelson}, R.~P., \& {Gammie}, C.~F. 2012, \apj, 746,
  110

\bibitem[{{Zucker} {et~al.}(2019){Zucker}, {Speagle}, {Schlafly}, {Green},
  {Finkbeiner}, {Goodman}, \& {Alves}}]{Zucker2019}
{Zucker}, C., {Speagle}, J.~S., {Schlafly}, E.~F., {et~al.} 2019, \apj, 879,
  125

\end{thebibliography}
